\documentclass[letterpaper]{article} % DO NOT CHANGE THIS
\usepackage[submission]{aaai25}  % DO NOT CHANGE THIS
\usepackage{aaai25}
\usepackage{times}  % DO NOT CHANGE THIS
\usepackage{helvet}  % DO NOT CHANGE THIS
\usepackage{courier}  % DO NOT CHANGE THIS
\usepackage[hyphens]{url}  % DO NOT CHANGE THIS
\usepackage{graphicx} % DO NOT CHANGE THIS
\usepackage{natbib}  % DO NOT CHANGE THIS AND DO NOT ADD ANY OPTIONS TO IT
\usepackage{caption} % DO NOT CHANGE THIS AND DO NOT ADD ANY OPTIONS TO IT
\usepackage{algorithm}
\usepackage{algorithmic}

\usepackage{graphicx}
\usepackage{amsmath}
\usepackage{amssymb}
\usepackage{multirow}
\usepackage{subcaption}
\usepackage{booktabs}
\usepackage{newfloat}
\usepackage{listings}
\DeclareCaptionStyle{ruled}{labelfont=normalfont,labelsep=colon,strut=off} % DO NOT CHANGE THIS
\floatstyle{ruled}
\newfloat{listing}{tb}{lst}{}
\floatname{listing}{Listing}
\title{Long-Tailed Backdoor Attack Using Dynamic Data Augmentation Operations}
\author{Lu Pang, Tao Sun, Weimin Lyu, Haibin Ling, Chao Chen\\
Stony Brook University\\
{\tt\small \{luppang,tao,welyu,hling\}@cs.stonybrook.edu, 
chao.chen.1@stonybrook.edu}}

\affiliations{
    \textsuperscript{\rm 1}Association for the Advancement of Artificial Intelligence\\
    1101 Pennsylvania Ave, NW Suite 300\\
    Washington, DC 20004 USA\\
    proceedings-questions@aaai.org
}

\usepackage{bibentry}
\begin{document}

\maketitle

\begin{abstract}
Recently, backdoor attack has become an increasing security threat to deep neural networks and drawn the attention of researchers. Backdoor attacks exploit vulnerabilities in third-party pretrained models during the training phase, enabling them to behave normally for clean samples and mispredict for samples with specific triggers.
Existing backdoor attacks mainly focus on balanced datasets. However, real-world datasets often follow long-tailed distributions. %, where head classes have more samples and tail classes contain fewer samples. 
In this paper, for the first time, we explore backdoor attack on such datasets. Specifically, we first analyze the influence of data imbalance on backdoor attack. %and observe that strong data augmentation can hinder the injection of stealthy triggers.
Based on our analysis, we propose an effective backdoor attack named \textit{Dynamic Data Augmentation Operation} (D$^2$AO). We design D$^2$AO selectors to select operations depending jointly on the class, sample type (clean vs. backdoored) and sample features. 
% Specifically, a clean augmentation selector is used to generate class-wise data augmentation for balancing clean classes, and a backdoored augmentation selector is used to generate weak instance-wise data augmentation for balancing backdoored samples and clean samples. 
Meanwhile, we develop a trigger generator to generate sample-specific triggers. 
Through simultaneous optimization of the backdoored model and trigger generator, guided by dynamic data augmentation operation selectors, we achieve significant advancements.
Extensive experiments demonstrate that our method can achieve the state-of-the-art attack performance while preserving the clean accuracy.
\end{abstract}


\section{Introduction}
\label{sec:intro}

Deep Neural Networks (DNN) have showcased remarkable achievements across various computer vision tasks, including image classification~\cite{he2016deep}, object detection~\cite{chen2023diffusiondet}, and visual object tracking~\cite{wei2023autoregressive, blatter2023efficient}. However, the state-of-the-art models like foundation models~\cite{radford2021learning, brown2020language} typically demand large scale training data and expensive computational resources. It is a common practice to outsource DNN training processes to third-party platforms or utilize pre-trained large-scale foundation models~\cite{radford2021learning, brown2020language} from such platforms. The opaque %nature of the 
training process raises concerns regarding trustworthiness and introduces security risks. 

Backdoor attack~\cite{barni2019sig, gu2019badnets, chen2017blended, doan2021lira, liu2020Refool} usually injects backdoor behaviors by attaching triggers to a portion of training samples and altering their labels into specified target labels. Backdoored models make correct predictions on clean samples but predict samples with triggers as target labels.
Despite a rich literature of backdoor attacks, the majority of existing methods assume a class-balanced data.
However, real-world data often exhibit a long-tailed distribution, where head classes have far more samples than tail classes~\cite{menon2020logitadj,ahn2022cuda}. For example, in autonomous driving scenarios, head classes typically consist of common objects like pedestrians and vehicles that have thousands of samples, while tail classes consist of rare objects like strollers and animals that only have tens of samples. Such imbalance issue inevitably introduces challenges to backdoor attack.

\begin{figure}[tp]
    \centering
    \includegraphics[width=0.95\columnwidth]{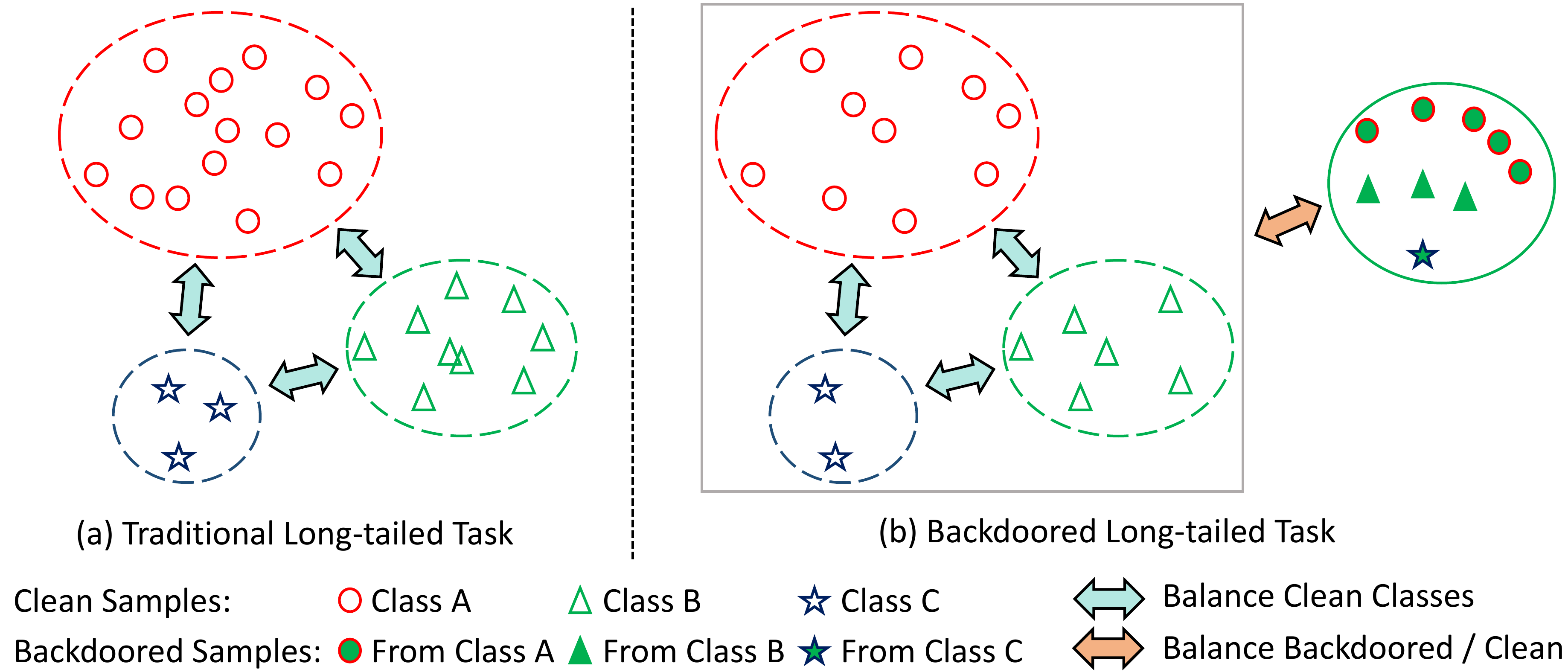}
    \caption{(a) Traditional long-tailed learning needs to balance different (clean) classes. (b) Long-tailed backdoor attack further needs to balance backdoored and clean samples, where backdoored samples can be viewed as a special class. (In the illustrations, backdoored samples are created from corresponding clean samples, and then relabeled as Class B.)}

    %\caption{(a) In traditional long-tailed learning, we only need to balance clean classes. (b) In long-tailed backdoor attack, we need to balance not only clean classes, but also backdoored samples and clean samples. (The target label is Class B, marked in green color.) %influence of strength of data augmentation (DA) on sample-specific backdoor attacks (IAB~\cite{nguyen2020IAB} and WaNet~\cite{nguyen2021wanet}). ASR means the classification accuracy on backdoored samples (higher indicating more effective).\lu{to be updated}} 
    \label{fig:intuition}
\end{figure}

% \begin{figure}
%     \centering
%     \includegraphics[width=0.9\columnwidth]{fig/ECCV2024-intuition.pdf}
%     \caption{(a) shows the difference of clean sample distributions between the original long-tailed dataset and the backdoored long-tailed dataset. (b) shows the influence of strength of data augmentation (DA) on sample-specific backdoor attacks (IAB~\cite{nguyen2020IAB} and WaNet~\cite{nguyen2021wanet}). ASR means the classification accuracy on backdoored samples (higher indicating more effective).
%     } 
%     \label{fig:intuition}
% \end{figure}

In this paper, for the first time, we study long tailed backdoor attack. We identify two key factors that hinder the attack efficacy: (1) the imbalanced sample sizes among different classes, and imbalance between clean and backdoored data;  (2) the data augmentation operations used for long-tail learning. 

\textbf{First}, as depicted in Fig.~\ref{fig:intuition}, backdoored samples can be viewed as a special class. The imbalance between clean samples and small-sized backdoored samples affects the attack learning. Besides, backdoored samples are created from different clean classes. This intra-imbalance also poses an obstacle to attacking tail classes successfully. 
\textbf{Second}, data augmentation is commonly used for long-tailed learning. While effective for improving classification performance, a strong augmentation can be detrimental to trigger generation in backdoor attack. A strong augmentation introduces large variations to the triggers, making it hard for the model to associate with and recognize triggers. In fact, existing works~\cite{borgnia2021strong, qiu2021deepsweep} have shown that strong augmentations neutralize backdoor attacks.

To solve the above challenges, we propose the first backdoor attack method for long-tailed datasets.
Our main idea is to apply novel Dynamic Data Augmentation Operations (D$^2$AO) on clean and backdoored samples. We design learnable selectors to dynamically determine augmentation operation types and strengths depending on the class, sample type (clean vs.~backdoored), and sample features. Specifically, we employ a Clean Selector (CS) to choose class-specific augmentation operations that ensures balance among clean classes. Then we employ a Backdoored Selector (BS) to generate weak instance-specific augmentation operations that maintains balance between clean and backdoored samples. 
Meanwhile, we train a trigger generator to adapt to data augmentation. The generated triggers are sufficiently resilient against weak data augmentation and can effectively flip model predictions into target labels during inference. 
%In order to balance sizes of clean classes and backdoor class, we propose a backdoor logits adjustment to adjust margin between backdoor samples and clean classes.
% To address above challenges, we propose a sample-specific backdoor attack with dynamic data augmentation operations for long-tailed visual recognition. First, we use a trigger generator to generate sample-specific triggers, which adapt to data augmentation operations of a specific sample. Second, we design a dynamic data augmentation operations selector to select suitable data augmentation operations for clean and backdoored samples respectively. 
% Since a backdoored long-tailed training dataset has fewer clean samples for each class than the original clean long-tailed dataset, we use a data augmentation operation selector to generate stronger data augmentation operations. In order to learn a sample-specific stealthy trigger, we design a backdoored data augmentation operation selector to choose weak data augmentation operations. Weak data augmentation operations can not destroy the learning of stealthy trigger patterns, and can improve the effectiveness of learned trigger patters simultaneously. 
We conduct experiments on two standard long-tailed benchmarks including CIFAR10-LT and CIFAR100-LT. Our method demonstrates state-of-the-art attack performance while preserving clean accuracy compared to other backdoor attack methods.

Our contributions are summarized as follows:
\begin{itemize}
    \vspace{-2mm}\item For the first time, we study backdoor attack for long-tailed visual recognition. We reveals key factors that hinder the attack efficacy under this practical setting. 
    %We analyze the impact of data augmentation on not only clean but also backdoored samples.
    \vspace{-1mm}
    \item We propose an effective long-tailed backdoor attack method through Dynamic Data Augmentation Operations (D$^2$AO). It chooses class-specific augmentations for clean images and generates instance-specific augmentations for backdoored images. 
    %It relies on the sample type (clean vs.~backdoored), the class and sample features. 
    \vspace{-1mm}
    \item We conduct extensive experiments on two long-tailed benchmarks, including CIFAR10-LT and CIFAR100-LT. Our method achieves state-of-the-art attack performances compared with other backdoor attacks.
    \vspace{-1mm}
\end{itemize}

\section{Related Work}
\label{sec:relatedwork}
\paragraph{Backdoor Attacks.}
Backdoor attacks~\cite{gu2019badnets, turner2019lc, barni2019sig, nguyen2020IAB, wang2022bppattack, nguyen2021wanet, liu2020Refool, liutrojaning, salem2022dynamic} introduce malicious behaviors by implanting specific triggers onto a small fraction of clean images, and subsequently altering their labels into a predefined target label. Models trained with such backdoored datasets exhibit accurate predictions for clean images, but  misclassify images containing triggers as the target label during inference. 
Most existing backdoor attacks focus on designing an effective trigger generator, which can be categorized into two main types: fixed trigger generators~\cite{gu2019badnets, turner2019lc, barni2019sig, liutrojaning, chen2017blended} and dynamic trigger generators~\cite{nguyen2020IAB, nguyen2021wanet, wang2022bppattack, liu2020Refool, doan2021lira}. Early works often employ fixed trigger generators to generate trigger patterns with consistent appearance and location across all samples, \textit{e.g.}, a checkerboard positioned at the bottom right corner of an image~\cite{gu2019badnets}. 
Such fixed triggers are not stealthy~\cite{nguyen2020IAB, nguyen2021wanet} and easily detected by humans. In constrast, recent works~\cite{nguyen2020IAB, nguyen2021wanet, wang2022bppattack, liu2020Refool, doan2021lira} use dynamic trigger generators to generate sample-specific trigger patterns that are stealthy and difficult to be detected. 
% The first proposed backdoor attacks~\cite{gu2019badnets, liutrojaning} usually use fixed trigger generators to generate the same trigger patterns for all backdoored samples. For example, BadNets~\cite{gu2019badnets} generates a checkerboard at the bottom right corner of an image and Liu et al.~\cite{liutrojaning} attach a color patch on the images. However, fixed trigger patterns are easy to be detected by human inspectors. 
% More backdoor attacks~\cite{nguyen2020IAB, wang2022bppattack, nguyen2021wanet, chen2017blended, doan2021lira, saha2020hidden} are developed to design trigger generators to generate more stealthy trigger patterns. Blended Attack~\cite{chen2017blended} generates backdoored images by blending clean images with the key trigger pattern. Salem et al.~\cite{salem2022dynamic} design a dynamic trigger generator to generate patch triggers with different colors and locations for different images.
% Input-aware backdoor attack (IAB)~\cite{nguyen2020IAB} and LIRA~\cite{doan2021lira} adopt U-Net to generate sample-specific triggers. Some backdoor attacks~\cite{wang2022bppattack, nguyen2021wanet} utilize image transformations as triger generation functions such as image quantization in BppAttack~\cite{wang2022bppattack} and image warping in WaNet~\cite{nguyen2021wanet}. In order to ensure backdoored images look natural, some works implicitly inject triggers in the physical space~\cite{wenger2021backdoor} or feature space~\cite{saha2020hidden}.
%
To further evade human inspection, clean-label backdoor attacks~\cite{turner2019lc, barni2019sig, liu2020Refool, shafahi2018poisonfrog} directly poison samples from the target label and do not modify ground-truth labels of backdoored samples. %This enables clean-label attacks to evade detection of erroneous labels by humans. 

Despite these efforts, current backdoor attacks all focus on balanced datasets. However, real-world datasets often exhibit long-tailed distributions. Head classes and tail classes exhibit different properties during training and inference, which hinder the attack efficacy of existing backdoor attacks.
In this paper, for the first time, we study backdoor attack for long-tailed visual recognition.
% In this paper, we will focus on attacking with dynamic triggers and clean-label attacks, due to their appealing features in practice.

\paragraph{Long-tailed Visual Recognition.}
% Models trained on long-tailed training datasets usually perform poorly on balanced testing datasets because models learn biases toward different classes. 
Methods for long-tailed visual recognition focus on balancing clean classes, and can be divided into two categories: re-weighting methods and re-sampling methods.

\textbf{Re-weighting methods}~\cite{alshammari2022long, cao2019learning, menon2020logitadj, yu2022re, guo2022learning, park2021IB, chen2023area, zhang2021FSR} increase the loss weights of tail classes to address the over-fitting issue of tail classes. 
Logits Adjustment (LA)~\cite{menon2020logitadj} incorporates offsets from class label frequencies into the logits of cross-entropy loss. Label-Distribution-Aware Margin (LDAM) loss~\cite{cao2019learning} enlarges margins of tail classes through margin regularization. %Cao et al.~\cite{cao2019learning} propose a label-distribution-aware margin (LDAM) loss, which enlarges margins of tail classes through margin regularization. %Alshammari et al.~\cite{alshammari2022long} balance the weights of loss by tuning weight decay and MaxNorm. To align imbalanced and balanced distributions, Guo et al.~\cite{guo2022learning} utilize optimal transport (OT) to guide learning of weights. 
AREA~\cite{chen2023area} calculates class weights inversely proportional to the effective statistical area of each class. 
Other methods~\cite{yu2022re, park2021IB, zhang2021FSR} focus on instance-level re-weighting by dynamically adjusting probabilities~\cite{yu2022re} of samples or identifying high-influence samples~\cite{park2021IB, zhang2021FSR}. %For example, Yu et al.~\cite{yu2022re} adopt instance-level re-weighting by dynamically adjusting probabilities of instances according to the learning speed of instances. IB~\cite{park2021IB} down-weights high-influence samples to mitigate the overfitting problem of decision boundaries. FSR~\cite{zhang2021FSR} utilizes the training history to discover meaningful samples for optimizing (increasing or decreasing) sample weights. 

These %re-weighting 
methods are designed for traditional long-tailed learning, and do not address the issue of imbalance between backdoored and clean samples. Our method can tackle this problem by applying appropriate data augmentation operations to clean and backdoored samples, respectively. % Our method improve the traditional logit adjustment and can address this problem.
%the distribution of backdoored dataset is different from the standard long-tailed distribution because the class of target label has more backdoored samples. Since it is difficult to determine the re-weighting ratios, we do not choose a re-weighting method to design backdoor attack method.

\textbf{Re-sampling methods}~\cite{ahn2022cuda, van2007experimental, sarafianos2018deep, mullick2019generative, kim2020m2m, li2021metasaug, park2022majority, chu2020feature} balance classes among imbalanced training datasets by either oversampling tail classes or undersampling head classes. Undersampling head classes~\cite{van2007experimental}, however, may discard valuable information. A simple oversampling technique is random oversampling, but it often leads to overfitting~\cite{sarafianos2018deep}. 
More sophisticated methods~\cite{ahn2022cuda, mullick2019generative, kim2020m2m, li2021metasaug, park2022majority} synthesize samples %for different classes 
to construct a balanced training dataset. For instance, Generative Adversarial Minority Oversampling (GAMO)~\cite{mullick2019generative} utilizes a convex generator to create new samples by combining samples from tail classes. 
MetaSAug~\cite{li2021metasaug} augments tail classes by translating deep semantic features along valuable semantic directions. 
M2m~\cite{kim2020m2m} and CMO~\cite{park2022majority} leverage sufficient information from head classes to augment tail classes.
%translates and leverages sufficient information of head classes to augment samples of tail classes. CMO~\cite{park2022majority} uses images of head classes as background images and paste images of tail classes on these background images. 
Similarly, Chu et al.~\cite{chu2020feature} augment feature space of tail classes with features learned from head classes. 
CUDA~\cite{ahn2022cuda} designs a class-wise data augmentation strategy to balance the performances of different classes.

Although these methods can improve performance for traditional long-tailed visual recognition, they lack robustness against backdoor attack. This is because strong data augmentation can impede the effectiveness of stealthy triggers~\cite{qiu2021deepsweep, borgnia2021strong}. %Another reason is that the number of clean samples in backdoored dataset is smaller than that of original long-tailed dataset. Applying existing data augmentation methods might not ensure a good classification accuracy for clean samples. 
Our method addresses this problem by designing a backdoored data augmentation operation selector to choose weak data augmentation operations, which are different from those selected by the clean data augmentation operation selector.
% two different data augmentation operations selectors to choose appropriate operations for clean and backdoored samples respectively.

\paragraph{Data Augmentation.}
Data augmentation addresses the overfitting problem due to the shortage of training data. For natural image dataset like CIFAR10~\cite{krizhevsky2009cifar10}, standard data augmentation operations such as image cropping and rotation are commonly used~\cite{krizhevsky2012imagenet, ciregan2012multi}. 
Recently, more advanced data augmentation methods are designed~\cite{cubuk2019autoaugment, cubuk2020randaugment, muller2021trivialaugment, cheung2021adaaug, suzuki2022teachaugment, hou2023learn}. 
AutoAugment~\cite{cubuk2019autoaugment} uses reinforcement learning to search best data augmentation operations. 
% RandAugment~\cite{cubuk2020randaugment} introduces a simplified search space. TrivialAugment~\cite{muller2021trivialaugment} also learns the minimal requirements for adaptive data augmentation algorithms. 
Some works~\cite{cubuk2020randaugment, muller2021trivialaugment} focus on improving efficiency by introducing a simplified search space~\cite{cubuk2020randaugment} or analyzing the minimal requirements~\cite{muller2021trivialaugment}.
TeachAugment~\cite{suzuki2022teachaugment} optimizes adversarial augmentation with ``teacher knowledge''. AdaAug~\cite{cheung2021adaaug} and MADAug~\cite{hou2023learn} propose class-wise or instance-wise adaptive data augmentation. 

Different from these works, we use dynamic data augmentation for both clean samples and backdoored samples. For clean samples, we apply class-specific data augmentation operations; while for backdoored samples, we apply instance-specific data augmentation operations. 

%only using one kind of data augmentation strategy, our method uses two kinds of data augmentation strategies. One strategy is to apply class-wise data augmentation operations to clean samples and the other one is to use instance-wise data augmentation operations to backdoored samples.
% Motivated by these works, we design data augmentation operations selectors for clean and backdoored samples. 

\section{Long-tailed Backdoor Attack}
\label{sec:definition}

\textbf{Threat Model.} We consider the same attack setting in previous works~\cite{saha2020hidden, nguyen2020IAB, nguyen2021wanet, turner2019lc} including the state-of-the-art attack called WaNet~\cite{nguyen2021wanet}. Attacker can totally control the training process, and change training dataset for backdoor injection. Then, the backdoored model will be delivered to the victim user. In the attack mode of image classification, a successfully backdoored model will predict correct label for clean images, but predict target label for images with triggers. 

\textbf{Problem Definition.}
Assuming there is a clean long-tailed training dataset $\mathcal{D}=\{(x_i, y_i)\}$ with input image $x_i \in \mathbb{R}^d$ and the corresponding label $y_i \in \{1, 2, ..., K\}$. $\mathcal{D}_k = \{(x, y)\in \mathcal{D}|y=k\}$ is the subset of $k$-th class. We assume that the number of samples decrease with class indexes, \textit{i.e.}, $N_{\rm max}=|\mathcal{D}_1| \geq |\mathcal{D}_2| \geq ... \geq |\mathcal{D}_K|=N_{\rm min}$. The imbalance ratio is defined as $IR = N_{\rm max} / N_{\rm min}$. 

Given a backdoor injection rate $\rho$, a subset $\mathcal{D}_m \subset \mathcal{D}$ is randomly selected with $|\mathcal{D}_m|=\rho * |\mathcal{D}|$. Then, the set of backdoored samples is constructed as $\mathcal{D}_b = \{(G'(x), \eta(y))|G'(x)=(1-\alpha)x+\alpha G(x), (x, y)\in\mathcal{D}_m\}$, where $\eta$ is a function mapping ground-truth label into target label, and $G$ is a backdoor trigger generator. $\alpha$ is the weight of blending generated trigger with the original image.
The objective function of long-tailed backdoor attack is:
\begin{equation}
{\scriptsize
\begin{aligned}
    \min_{\theta, G} \Big\{\mathbb{E}_{(x, y)\thicksim \mathcal{D}\backslash\mathcal{D}_m}\mathcal{L}_c(f_\theta(x), y) +\mathbb{E}_{(x, y)\thicksim \mathcal{D}_m}\mathcal{L}_b(f_\theta(G'(x)), \eta(y))\Big\}
 \end{aligned}
\label{equ:backdoor_attack}
}
\end{equation}
where $f_{\theta}$ is a classification network parameterized by $\theta$. $\mathcal{L}_c$ and $\mathcal{L}_b$ are classification loss functions for clean samples and backdoor samples, respectively. 

\begin{figure*}[tp]
    \centering
    \includegraphics[width=0.9\linewidth]{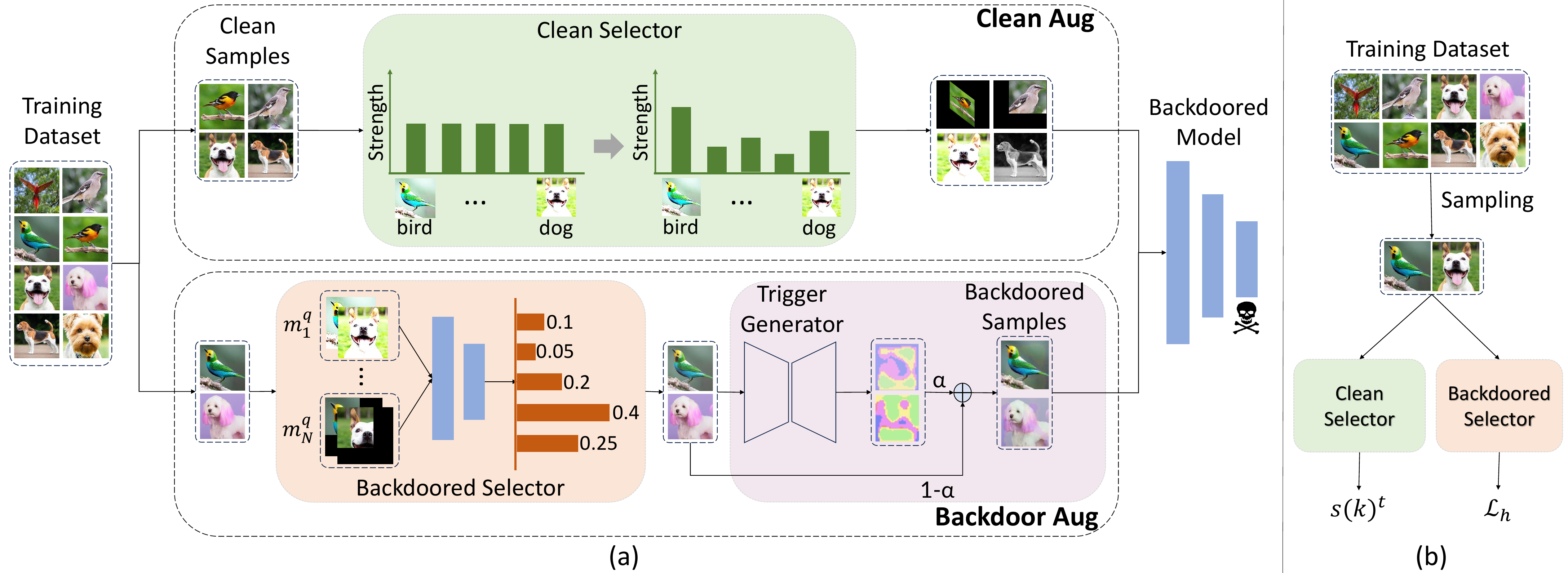}
    \caption{Framework of our method including (a) Backdoored Model Training and (b) Operation Selectors Training. %Two data augmentation operation selectors including clean operation selector and backdoored operation selector are designed to guide the data augmentation of samples. 
    At stage (a), Clean Selector chooses class-wise augmentation operations for clean samples. Backdoored Selector calculates a probability distribution over $N$ data augmentation operations with fixed strength $q$, and chooses instance-wise operations based on the probabilities. Trigger Generator adapts to data augmentation of backdoored samples for generating effective trigger patterns. %$\alpha$ is the blending strength of trigger. 
    Trigger generator and backdoored model are optimized simultaneously. 
    At stage (b), clean and backdoored selectors are updated based on strength score $s(k)^t$ and the proposed loss $\mathcal{L}_h$.
    At each epoch, (b) is first conducted and then (a) is conducted.
    }
    \label{fig:architecture}
\end{figure*}

\textbf{Problem Analysis.} Since $\mathcal{D}_m$ is randomly selected from $\mathcal{D}$ with the same probability for every sample, it contains more samples from head classes and fewer samples from tail classes. The consequent imbalance in the constructed backdoor samples makes it difficult to effectively flip tail classes into target label, which thereby increases the difficulty of optimizing $\mathcal{L}_b$. Additionally, the imbalance between clean samples and backdoored samples affects the optimization of both $\mathcal{L}_b$ and $\mathcal{L}_c$. As illustrated in Fig.~\ref{fig:intuition}, when the target label is in head or medium classes, the number of backdoored samples and clean samples with the target label are comparable. This makes it challenging for the model to classify two distinct types of samples into the same label. When the target label is in tail classes, the class with the target label becomes dominated by backdoored samples. In both cases, there exists a trade-off between optimization of $\mathcal{L}_b$ and $\mathcal{L}_c$. 
Besides, the imbalanced sizes among clean classes also hinder the optimization of $\mathcal{L}_c$. 
The above analysis motivates us to adopt different data augmentation operations for clean and backdoored samples, respectively.

\section{Method}
\label{sec:method}

\subsection{Overview}

To successfully train a backdoored model on long-tailed datasets, we propose a method that finds Dynamic Data Augmentation Operations (D$^2$AO) for clean and backdoored samples simultaneously. 
% By applying selected proper data augmentation operations, we can improve attack performance of trained backdoored model and meanwhile maintain the clean accuracy.
Shown in Figure~\ref{fig:architecture}, our framework includes backdoored model training and operation selectors training. The two training procedures are conducted alternatively. At each epoch, we first train operation selectors and then train backdoored model.

At model training stage, the Clean Selector can choose class-specific data augmentation operations with varying augmentation strengths for clean samples. For backdoored samples, the Backdoored Selector chooses sample-specific data augmentation operations based on predicted probabilities over these operations. Augmented images are feed into a trigger generator to generate perturbation triggers, which are attached on the original images to construct backdoored images. Both augmented clean images and backdoored images are utilized to train the backdoored model.

At operation selectors training stage, we sample a fraction of images from training dataset to update the data augmentation operation selectors.  
For the Clean Selector, we increase data augmentation strength if the model can classify weakly augmented images with a good performance. For the Backdoored Selector, we learn a network to predict the probabilities of data augmentation operations by leveraging the classification accuracy.

\subsection{Backdoored Model Training}

\textbf{Data Augmentation for Clean Samples.}
Since clean samples is class-imbalanced, we design a Clean Selector (CS) to select class-specific data augmentation operations for balancing different clean classes. 
Assume there are $N$ available data augmentation operators (e.g. Flip and Rotation). A data augmentation operator with a specific augmentation strength level $s$ is considered as a data augmentation operation $m^s$. The set of data augmentation operations is defined as $\mathcal{M}=\{m_i^s|i\in\{1, 2, ..., N\}, s\in\{1, 2, ..., s_{\rm max}\}\}$, where $s_{\rm max}$ is the maximum level of data augmentation strength. 
For class $k$, we introduce a learned parameter $s(k)$ to indicate the strength of data augmentation applied to samples from class $k$. 
We will randomly choose $n(k)$ data augmentation operations from the data augmentation operation set $\mathcal{M}_{s(k)}=\{m^j|m^j\in\mathcal{M}, j=s(k)\}$. Following a state-of-the-art data augmentation based long-tailed method CUDA~\cite{ahn2022cuda}, we set $n(k) = s(k)$ in our experiments. We will discuss the update of $s(k)$ in Sec~\ref{Sec:DA_training}. 

% We define $\tau_k = m_{i^k}^{s(i^k)}\circ...\circ m_2^{s(2)}\circ m_1^{s(1)}$ as a series of data augmentation operations for clean samples from class $k$, and $i^k$ means the number of data augmentation operations for clean samples from class $k$. 

\textbf{Data Augmentation for Backdoored Samples.}
To balance clean samples and backdoored samples, we adopt different data augmentation operations for clean and backdoored samples separately. Since backdoored samples can come from all classes and each backdoored sample is critical for attack performance, we design a Backdoored Selector (BS) to choose sample-specific data augmentations. Previous works~\cite{qiu2021deepsweep, borgnia2021strong} have shown that strong augmentations can mitigate backdoor attack to some extent. Therefore, our designed backdoored selector should choose weak sample-wise operations. Specifically, we utilize a trained data augmentation selection network $h$ to predict probabilities which indicate the strength of data augmentations. High probabilities means weak augmentations while low probabilities correspond to strong augmentations. 
Given a fixed data augmentation strength $q$, we use the trained data augmentation selection network $h$ to predict probabilities of all $N$ data augmentation operations in $\mathcal{M}_{q}=\{m^j|m^j\in\mathcal{M}, j=q\}$. Based on these predicted probabilities, $n(q)$ data augmentation operations are randomly chosen from $\mathcal{M}_{q}$ to apply to each backdoored sample. Similar to clean augmentation, we set $n(q) = q$ in our experiments. The update of data augmentation selection network $h$ will be discussed in Sec~\ref{Sec:DA_training}.

% We also define $\tau_b = m_{i^b}^{s(i^b)}\circ...\circ m_2^{s(2)}\circ m_1^{s(1)}$ as a series of data augmentation operations for backdoored samples. 
% Therefore, the optimization objective (Eq.~\ref{equ:backdoor_attack}) can be transformed into:
% \begin{equation}
% \begin{aligned}
%     \min_{\theta, G} &\Big\{\mathbb{E}_{(x, y)\thicksim \mathcal{D}\backslash\mathcal{D}_m}\mathcal{L}(f_\theta(\tau_y(x)), y) \\
%    & +\mathbb{E}_{(x, y)\thicksim \mathcal{D}_m}\mathcal{L}(f_\theta(G(\tau_b(x))), \eta(y))\Big\} 
%     \end{aligned}
%     \label{equ:da_attack}
% \end{equation}

\textbf{Sample-Specific Trigger Generator.}
To evade human inspection and ensure the stealthiness of triggers, we utilize an auto-encoder based trigger generator $G$ to generate sample-specific global perturbation triggers conditioned on specific clean images. The generated trigger is then attached with blending strength $\alpha$ on the clean image to construct a backdoored image.
To ensure to generate sample-specific triggers, we follow IAB~\cite{nguyen2020IAB} to enforce trigger diversity by using a diversity loss:
\begin{equation}
    \mathcal{L}_{div} = \mathbb{E}_{(x, y)\thicksim\mathcal{D}_m}\frac{\|x - x'\|}{\|G(BS(x)) - G(BS(x'))\|}
\end{equation}
where for each $x$, $x'$ is randomly chosen from $\mathcal{D}\backslash\mathcal{D}_m$.

We attach triggers after conducting data augmentation on samples from $\mathcal{D}_m$ to ensure that trigger generator $G$ adapts to data augmentation operations. Both augmented clean and backdoored samples are used to train the backdoored model according to the objective function in Eq.~\ref{equ:backdoor_attack}.

% The objective (Eq.~\ref{equ:da_attack}) is transformed as:
% \begin{equation}
% \begin{aligned}
%     \min_{\theta, G} &\Big\{\mathbb{E}_{(x, y)\thicksim \mathcal{D}\backslash\mathcal{D}_m}\mathcal{L}(f_\theta(\tau_y(x)), y) \\
%    & +\mathbb{E}_{(x, y)\thicksim \mathcal{D}_m}\mathcal{L}(f_\theta(G(\tau_b(x))), \eta(y)) + \lambda\mathcal{L}_{div}\Big\} 
%     \end{aligned}
%     \label{equ:da_div_attack}
% \end{equation}
% where $\lambda$ controls the weight of $\mathcal{L}_{div}$. 

\subsection{Data Augmentation Operation Selectors Training}
\label{Sec:DA_training}

% The number of per-class clean samples usually decreases in a backdoorded long-tailed dataset (shown in Figure~\ref{fig:intuition}). Compared to the original long-tailed dataset, it is more difficult to learn a model that achieves a better average accuracy for clean samples. With fewer clean samples, strong data augmentation operations tend to be applied to augment feature space. 
% However, strong data augmentation can impede the learning of trigger generator. The reason is that slight perturbation generated by trigger generator can not flip ground-truth labels of strongly augmented images.

% To balance clean classes and backdoored samples, we propose to different data augmentation operations selectors to choose appropriate data augmentation operations for clean and backdoored samples respectively. 
The data augmentation operation selectors and backdoored model are trained alternatively. 
At the $t$-th epoch, we first sample a fraction of images from each class to construct a temporary dataset $\mathcal{D}_t$ for updating the operation selectors. Then we train the classifier $f_\theta$ and the trigger generator $G$ guided by fixed selectors at the $t$-th epoch.
The following describes how the selectors are updated at each epoch.

\textbf{Clean Selector Training.}
The Clean Selector is designed to determine how to choose appropriate data augmentation operations for clean samples. The core idea is to first sufficiently learn representations of weakly augmented images, and then learn representations of strongly augmented images. We denote $s(k)^t$ as the augmentation strength score of class $k$ at the $t$-th epoch.
Thus, we update $s(k)^t$ for class $k$ based on $\mathcal{D}_t$ as follows:
\begin{equation}  
    s(k)^t = \left\{\begin{array}{cl}
         s(k)^{t-1}+1& \quad \mbox{if } acc(k)>\gamma\\
         s(k)^{t-1}-1& \quad \mbox{otherwise}
    \end{array}\right.
\end{equation}
where $acc(k)$ denotes the average accuracy of class $k$ in $\mathcal{D}_t$. When calculating $acc(k)$, images of class $k$ are augmented by randomly selecting $s(k)^{t-1}$ data augmentation operations from the operation set 
$\mathcal{M}_{s(k)^{t-1}}=\{m^j|m^j\in\mathcal{M}, j=s(k)^{t-1}\}$. Once $s(k)^t$ for the $t$-th epoch is determined, we utilize the updated Clean Selector to augment images in $\mathcal{D}\backslash\mathcal{D}_m$ for training the backdoored model $f_\theta$.

\textbf{Backdoored Selector Training.}
The Backdoored Selector is employed to choose proper data augmentation operations for backdoored samples. Given a fixed strength $q$, the core idea is to determine which operations are weak data augmentation operations for each input image $x$. We use an operation selector network $h$ to predict probabilities of all operations in the operations set with the fixed strength $q$. For an input image $x$, the network outputs high probabilities for weak data augmentation operations and low probabilities for strong data augmentation operations. 

In comparison to strongly augmented samples, weakly augmented samples are easy to achieve higher classification accuracy. Therefore, we can optimize an operation selector network $h$ by using a classification loss, e.g., cross entropy loss. 
% We use a operation selector network $h$ (e.g. two fc layers) to estimate the probabilities of each data augmentation operations with fixed strength $q$ on the sampled dataset $\mathcal{D}_t$. $q$ is a hyperparameter. 
For a given input image $x$ and a strength $q$, we augment $x$ with $N$ data augmentation operations ($m_1^q, m_2^q, ..., m_N^q$) respectively. $N$ augmented images are then fed into a feature generator $f_\theta^{*}$ to get $N$ features. The aggregated feature of $x$ is obtained by adding $N$ features weighted by output probabilities of $h$ (rescaled with temperature $T$). Then, the aggregated feature is input into a classifier $f'$ to obtain classification logits. The final optimization objective of $h$ is:
\begin{equation}
{\scriptsize
\begin{aligned}
    \mathcal{L}_h = \mathbb{E}_{(x, y)\thicksim\mathcal{D}_t} \ l_{ce} \Bigg(f'\Big(\sum_{i=1}^N \text{Softmax}\big(h(f_\theta^{*}(x)), T\big)[i]\cdot f_\theta^{*}\big(m_i^q(x)\big) \Big), y \Bigg)
\end{aligned}
}
\end{equation}
where $l_{ce}$ is the cross entropy loss. Through the loss, $h$ is encouraged to output high probabilities for data augmentation operations that are less likely to change the ground-truth labels. Consequently, we obtain probabilities that indicate which data augmentations are weak for trigger learning.

\begin{table*}[!tp]
\footnotesize
% \small
    \centering
    \scalebox{0.95}{
    \renewcommand{\tabcolsep}{0.1cm}
    \begin{tabular}{c|c|c|c|c|c|c|c|c|c|c|c|c|c|c}
\toprule
\multirow{2}{*}[-0.4em]{Metric} 
% & \multirow{2}{*}{\begin{tabular}[c]{@{}c@{}}\textbf{Backdoor}\\ \textbf{Attacks}\end{tabular}} 
& \multirow{2}{*}[-0.4em]{Attack}
& \multicolumn{4}{c|}{Target Label: Many} & \multicolumn{4}{c|}{Target Label: Medium} & \multicolumn{4}{c|}{Target Label: Few} 
& \multirow{2}{*}[-0.4em]{Avg}\\ 
\cmidrule{3-14} 
& & Many & Med. & Few & All 
& Many & Med. & Few & All 
& Many & Med. & Few & All 
\\ 
\midrule
\multirow{5}{*}[-0.3em]{ACC} &
% \textbf{BN}& 90.89 & 67.78 & 54.67 & 69.51 & 90.56 & 70.00 & 56.50 & 70.69 & 90.92 & 68.00 & 57.31 & 70.58 &70.29\\
% \textbf{BN+C}& 92.89 & 70.67 & 60.75 & 73.38 & 92.89 & 75.00 & 62.75 & 75.54 & 92.92 & 73.67 & 62.31 & 74.93 &74.65\\
% \textbf{Bl}& 92.67 & 74.78 & 60.08 & 74.39 & 92.78 & 75.33 & 59.75 & 74.33 & 92.83 & 74.58 & 60.62 & 74.36 &74.36\\
% \textbf{Bl+C}& 92.67 & 74.78 & 60.08 & 74.39 & 92.78 & 75.33 & 59.75 & 74.33 & 92.83 & 74.58 & 60.62 & 74.36 &74.36\\
% \textbf{IAB}& 92.81 & 73.58 & 58.54 & 73.33 & 92.50 & 74.24 & 57.33 & 72.95 & 92.95 & 74.07 & 58.20 & 73.39 &73.24\\
IAB & 92.79 & 75.63 & 65.34 & 76.66 & 92.76 & 75.22 & 63.81 & 75.92 & 92.73 & 75.92 & 64.86 & 76.54 &76.39\\
% \textbf{LC}& 91.56 & 75.56 & 61.25 & 74.70 & 93.44 & 72.11 & 61.42 & 74.21 & 92.92 & 74.50 & 58.19 & 73.47 &74.06\\
& LC & 94.00 & 76.78 & 68.00 & 78.38 & 94.67 & 74.22 & 68.00 & 77.86 & 94.58 & 76.83 & 66.31 & 77.94 &78.05\\
% \textbf{SIG}& 92.00 & 74.78 & 62.67 & 75.13 & 93.33 & 70.22 & 61.58 & 73.70 & 93.25 & 74.08 & 57.63 & 73.27 &73.95\\
& SIG & 92.00 & 74.78 & 62.67 & 75.13 & 93.33 & 70.22 & 61.58 & 73.70 & 93.25 & 74.08 & 57.63 & 73.27 &73.95\\
% \textbf{WN}& 91.80 & 66.31 & 53.80 & 68.95 & 91.99 & 67.77 & 54.75 & 69.83 & 92.18 & 67.90 & 53.76 & 69.53 &69.45\\
& WN & 92.67 & 71.29 & 62.97 & 74.37 & 92.72 & 73.49 & 63.15 & 75.12 & 92.45 & 74.97 & 65.44 & 76.40 &75.41\\
\cmidrule{2-15}
& Ours & 93.83 & 73.32 & 62.08 & 74.98 & 92.76 & 74.60 & 65.97 & 76.60 & 93.08 & 75.85 & 64.10 & 76.32 &76.00\\

\midrule
\midrule
\multirow{5}{*}[-0.3em]{ASR} 
% \textbf{BN}& 99.33 & 97.00 & 95.00 & 96.57 & 96.11 & 93.67 & 92.25 & 93.82 & 95.25 & 92.75 & 93.00 & 93.77 &94.62\\
% \textbf{BN+C}& 100.00 & 99.89 & 98.83 & 99.43 & 100.00 & 99.83 & 99.17 & 99.54 & 100.00 & 99.75 & 99.33 & 99.65 &99.55\\
% \textbf{Bl}& 99.67 & 99.44 & 96.83 & 98.26 & 99.67 & 99.67 & 96.92 & 98.36 & 99.67 & 100.00 & 97.83 & 99.01 &98.59\\
% \textbf{Bl+C}& 99.67 & 99.44 & 96.83 & 98.26 & 99.67 & 99.67 & 96.92 & 98.36 & 99.67 & 100.00 & 97.83 & 99.01 &98.59\\
% \textbf{IAB}& 96.80 & 97.02 & 95.97 & 96.51 & 96.79 & 95.72 & 96.76 & 96.54 & 96.36 & 96.08 & 96.74 & 96.39 &96.47\\
& IAB & 94.42 & 94.51 & 92.26 & 93.49 & 92.73 & 94.23 & 91.08 & 92.33 & 94.28 & 94.10 & 93.36 & 93.91 &93.31\\
% \textbf{LC}& 93.00 & 96.78 & 97.17 & 96.13 & 77.89 & 94.17 & 89.75 & 86.71 & 14.50 & 32.58 & 22.17 & 23.13 &64.10\\
& LC & 80.33 & 88.11 & 88.83 & 86.67 & 89.56 & \textbf{99.50} & 96.42 & 94.81 & 26.17 & 44.25 & 30.67 & 33.71 &67.93\\
% \textbf{SIG}& 89.83 & 89.56 & 92.33 & 90.81 & 79.44 & 92.67 & 84.58 & 84.60 & 43.83 & 40.50 & 39.92 & 41.43 &69.19\\
& SIG & 89.83 & 89.56 & 92.33 & 90.81 & 79.44 & 92.67 & 84.58 & 84.60 & 43.83 & 40.50 & 39.92 & 41.43 &69.19\\
% \textbf{WN}& 95.32 & 98.10 & 97.44 & 97.19 & 95.58 & 97.95 & 97.58 & 97.00 & 95.33 & 98.22 & 97.88 & 97.15 &97.11\\
& WN & 93.37 & 95.37 & 95.65 & 95.05 & 94.04 & 97.75 & 96.18 & 95.82 & 95.12 & \textbf{97.48} & \textbf{97.17} & 96.59 &95.90\\
\cmidrule{2-15}
& Ours & \textbf{97.25} & \textbf{96.93} & \textbf{96.25} & \textbf{96.70} & \textbf{98.30} & 97.42 & \textbf{96.93} & \textbf{97.49} & \textbf{97.84} & 96.52 & 96.29 & \textbf{96.88} & \textbf{97.01}\\
\bottomrule
\end{tabular}
    }
    \vspace{-2mm}
\caption{Comparison results on CIFAR10-LT.}
\label{tab:cifar10-c}
\end{table*}

\begin{table*}[!t]
\footnotesize
% \small
    \centering
    \scalebox{0.95}{
    \renewcommand{\tabcolsep}{0.1cm}
    \begin{tabular}{c|c|c|c|c|c|c|c|c|c|c|c|c|c|c}
\toprule
\multirow{2}{*}[-0.4em]{Metric} 
% & \multirow{2}{*}{\begin{tabular}[c]{@{}c@{}}\textbf{Backdoor}\\ \textbf{Attacks}\end{tabular}} 
& \multirow{2}{*}[-0.4em]{Attack}
& \multicolumn{4}{c|}{Target Label: Many} & \multicolumn{4}{c|}{Target Label: Medium} & \multicolumn{4}{c|}{Target Label: Few} 
& \multirow{2}{*}[-0.4em]{Avg}\\ 
\cmidrule{3-14} 
& & Many & Med. & Few & All 
& Many & Med. & Few & All 
& Many & Med. & Few & All 
\\ 
\midrule
\multirow{5}{*}{ACC} 
% \textbf{BN}& 64.42 & 49.68 & 33.74 & 49.13 & 63.70 & 49.83 & 34.18 & 49.08 & 64.55 & 50.71 & 35.19 & 50.00 &49.41\\
% \textbf{BN+C}& 67.93 & 55.36 & 40.85 & 54.58 & 68.56 & 55.78 & 40.25 & 54.72 & 68.13 & 54.99 & 40.87 & 54.53 &54.61\\
% \textbf{Bl}& 66.51 & 52.35 & 36.21 & 51.54 & 66.71 & 51.04 & 36.01 & 51.10 & 67.34 & 52.51 & 36.20 & 51.86 &51.50\\
% \textbf{Bl+C}& 71.65 & 57.99 & 43.17 & 57.46 & 71.82 & 57.88 & 43.06 & 57.44 & 71.97 & 58.18 & 42.75 & 57.48 &57.46\\
% \textbf{IAB}& 66.77 & 52.68 & 37.51 & 52.17 & 66.27 & 52.29 & 38.01 & 52.05 & 66.85 & 52.60 & 37.59 & 52.20 &52.14\\
& IAB & 70.64 & 57.05 & 40.79 & 56.00 & 70.32 & 57.10 & 40.82 & 55.93 & 70.92 & 56.85 & 41.09 & 56.13 &56.02\\
% \textbf{LC}& 67.54 & 53.64 & 37.76 & 52.83 & 68.25 & 53.17 & 38.03 & 53.00 & 67.31 & 52.94 & 37.10 & 52.30 &52.70\\
& LC & 70.35 & 57.48 & 42.76 & 56.72 & 72.14 & 58.42 & 44.19 & 58.11 & 69.07 & 58.28 & 42.50 & 56.48 &57.10\\
% \textbf{SIG}& 66.95 & 54.06 & 38.12 & 52.90 & 68.27 & 52.56 & 37.74 & 52.70 & 68.73 & 53.59 & 37.10 & 52.98 &52.86\\
& SIG & 71.54 & 59.06 & 45.01 & 58.40 & 69.16 & 55.44 & 39.78 & 54.65 & 71.51 & 58.64 & 41.21 & 56.96 &56.67\\
% \textbf{WN}& 67.60 & 53.81 & 38.84 & 53.27 & 68.59 & 53.89 & 38.44 & 53.49 & 67.57 & 54.20 & 38.19 & 53.17 &53.31\\
& WN & 71.59 & 58.88 & 45.22 & 58.43 & 70.92 & 59.18 & 45.23 & 58.31 & 71.41 & 58.96 & 45.84 & 58.61 &58.45\\
\cmidrule{2-15}
& Ours & 70.83 & 56.58 & 41.25 & 56.07 & 70.52 & 57.07 & 41.56 & 56.23 & 71.43 & 56.27 & 40.35 & 55.86 &56.05\\
\midrule
\midrule
\multirow{5}{*}{ASR} 
% \textbf{BN}& 99.51 & 99.08 & 99.71 & 99.43 & 99.58 & 99.21 & 99.75 & 99.52 & 99.34 & 98.94 & 99.53 & 99.27 &99.41\\
% \textbf{BN+C}& 99.92 & 99.50 & 99.76 & 99.72 & 99.95 & 99.73 & 99.97 & 99.89 & 99.82 & 99.48 & 99.78 & 99.69 &99.77\\
% \textbf{Bl}& 99.98 & 99.93 & 99.99 & 99.97 & 100.00 & 99.93 & 99.99 & 99.97 & 99.98 & 99.90 & 100.00 & 99.96 &99.97\\
% \textbf{Bl+C}& 99.83 & 99.64 & 99.80 & 99.76 & 99.80 & 99.72 & 99.80 & 99.77 & 99.82 & 99.59 & 99.87 & 99.76 &99.76\\
% \textbf{IAB}& 98.66 & 99.12 & 99.30 & 99.03 & 98.93 & 99.43 & 99.53 & 99.30 & 98.91 & 99.10 & 99.28 & 99.10 &99.14\\
& IAB & 97.92 & 98.07 & 98.26 & 98.09 & 98.09 & 98.08 & 98.42 & 98.20 & 97.58 & 97.24 & 98.16 & 97.66 &97.98\\
% \textbf{LC}& 47.63 & 50.21 & 57.63 & 51.93 & 48.73 & 50.24 & 58.63 & 52.62 & 14.94 & 20.93 & 20.87 & 18.91 &40.93\\
& LC & 38.13 & 40.05 & 44.51 & 40.96 & 43.96 & 43.95 & 51.67 & 46.60 & 3.34 & 4.89 & 5.43 & 4.56 &30.45\\
% \textbf{SIG}& 88.18 & 88.80 & 90.41 & 89.15 & 75.05 & 75.97 & 77.58 & 76.22 & 75.85 & 79.88 & 78.89 & 78.21 &81.16\\
& SIG & 94.63 & 95.11 & 96.80 & 95.54 & 90.84 & 91.59 & 94.04 & 92.18 & 89.05 & 89.43 & 91.83 & 90.10 &92.58\\
% \textbf{WN}& 96.34 & 97.50 & 96.09 & 96.64 & 97.00 & 98.17 & 96.94 & 97.36 & 97.03 & 97.77 & 96.94 & 97.25 &97.08\\
& WN & 91.26 & 94.42 & 91.96 & 92.55 & 92.04 & 95.22 & 92.80 & 93.33 & 91.68 & 94.76 & 91.69 & 92.71 &92.86\\
\cmidrule{2-15}
& Ours & \textbf{98.60} & \textbf{98.75} & \textbf{98.82} & \textbf{98.72} & \textbf{98.83} & \textbf{98.71} & \textbf{98.85} & \textbf{98.80} & \textbf{98.27} & \textbf{98.23} & \textbf{98.67} & \textbf{98.39} &\textbf{98.64}\\
\bottomrule
\end{tabular}
    }
    \vspace{-2mm}
    \caption{Comparison results on CIFAR100-LT.}
    \label{tab:cifar100-c}
    
\end{table*}

\section{Experiments}
\label{sec:experiments}

\subsection{Experiments Details}
\textbf{Datasets.}
We conduct experiments on two long-tailed benchmarks called CIFAR10-LT and CIFAR100-LT, which are constructed from balanced CIFAR10~\cite{krizhevsky2009cifar10} and CIFAR100~\cite{krizhevsky2009cifar10}.
The Imbalance Ratios (IR) of CIFAR10 and CIFAR100 are set as 50 and 10, respectively. 

\textbf{Evaluation Metrics.}
The evaluation metrics consist of clean accuracy (ACC) and attack success rate (ASR). 
% ACC is the classification accuracy on clean samples, reflecting the classification ability of model on clean samples. ASR is the classification accuracy on backdoored images, indicating the attack ability of model on backdoored images. 
ACC is the classification accuracy on clean samples and ASR is the classification accuracy on backdoored images.
We further follow traditional long-tailed visual recognition to divide all classes into three groups: ``Many'', ``Medium (Med.)'' and ``Few''. The classes are sorted descending based on the number of class samples. ``Many'' consists of classes with first 1/3 classes. ''Medium (Med.)'' comprises those of second 1/3 classes. ``Few'' composes of last 1/3 classes. 
We use three types of target labels to attack models. For each kind of target label, we compute the average ACC of three groups, and the average ASR for backdoored samples from three kinds of source labels. We also compute the average ACC and ASR across ``ALL'' classes.

\textbf{Backdoor Attack Methods.}
Considering the practical application, we compare with four state-of-the-art stealthy backdoor attacks including Label-consistent backdoor attack (LC)~\cite{turner2019lc}, Sinusoidal signal backdoor attack (SIG)~\cite{barni2019sig}, Input-aware backdoor attack (IAB)~\cite{nguyen2020IAB} and Warp-based backdoor attack (WaNet)~\cite{nguyen2021wanet}. 
IAB~\cite{nguyen2020IAB} and WaNet~\cite{nguyen2021wanet} represent dirty-label attacks with dynamic trigger generators. LC~\cite{turner2019lc} and SIG~\cite{barni2019sig} are two classic clean-label attacks.
% Compared to dirty-label attacks, clean-label attacks are more stealthy. Therefore, we also compare with two classic clean-label attacks including LC~\cite{turner2019lc} and SIG~\cite{barni2019sig}. % More experiments details about four backdoor attacks are shown in supplementary.

\textbf{Long-tailed Visual Recognition Methods.}
We choose a state-of-the-art data augmentation based re-sampling method called CUDA~\cite{ahn2022cuda} and a classic re-weighting method called Logit Adjustment (LA)~\cite{menon2020logitadj}.  The two methods are integrated with backdoor attack methods for comparing performance.

\textbf{Training Settings.} 
The architecture of classification model $f_\theta$ is ResNet18~\cite{he2016deep}, and the architecture of trigger generator $G$ is an auto-encoder similar to that in IAB~\cite{nguyen2020IAB}. We optimize classification model $f_\theta$ using Stochastic Gradient Descent (SGD) optimizer with a momentum of $0.9$ and a weight decay of $0.0005$. The learning rate is set as $0.01$. We optimize trigger generator $G$ using Adam optimizer with same learning rate $0.01$. The backdoored operation selector network $h$ consists of a FC layer. The network $h$ is optimized using Adam optimizer with learning rate $0.01$. In the implementations, $f_\theta^{*}$ is the last FC layer in $f_\theta$ and $f_\theta^{*}$ is the rest part of $f_\theta$ excluding last FC layer ($f_\theta^{*}$). The setting of other hyperparameters areas follows: $\lambda_{div}=0.01$, $q=1$ and $\alpha=0.1$. We use data augmentation with probability of $0.5$ for clean samples. We conduct experiments on a A6000 GPU (48G memory).

\subsection{Results compared with other attacks}
We compare with four state-of-the-art stealthy backdoor attacks including LC~\cite{turner2019lc}, SIG~\cite{barni2019sig}, IAB~\cite{nguyen2020IAB} and WaNet~\cite{nguyen2021wanet}. To ensure fair comparison, we integrate these four attacks with a state-of-the-art data augmentation based long-tailed method called CUDA~\cite{ahn2022cuda}. CUDA~\cite{ahn2022cuda} utilizes class-wise data augmentation strategies based on the features of clean samples. Therefore, a backdoored sample will be augmented using data augmentation operations specific to its original class when integrating backdoor attacks with CUDA~\cite{ahn2022cuda}. %In constrast, our method augments a backdoored sample using dynamically selected data augmentation operations based on its individual features. 
We also conduct experiments combining all data augmentation based methods, including our method, with a classic re-weighting long-tailed method called Logits Adjustment (LA)~\cite{menon2020logitadj}. LA adjusts decision boundaries between head classes and tail classes by adjusting logits. 
We use all-to-one attacks and compute average results on three kinds of target labels: ``Many'', ``Medium (Med.)'' and ``Few''. For each target label category , we compute the average results across three groups of classes. We also compute average ACC and ASR across ``ALL'' classes.

Table~\ref{tab:cifar10-c} presents the results on CIFAR10-LT compared with other backdoor attacks integrated with CUDA~\cite{ahn2022cuda}, and Table~\ref{tab:cifar100-c} shows the results on CIFAR100-LT. We can observe that our method achieves state-of-the-art attack performance (ASR) in most cases on CIFAR10-LT and all cases on CIFAR100-LT while preserving the clean accuracy (ACC). This demonstrates that our backdoored selector can generate suitable data augmentation operations for backdoored samples. 
% For sample-specific attacks (IAB~\cite{nguyen2020IAB} and WaNet~\cite{nguyen2021wanet}), using a data augmentation long-tailed method (CUDA~\cite{ahn2022cuda}) can increase ACC but decrease ASR. This indicates that some data augmentation operations can hinder the learning of trigger generator. 

In contrast to the performance when the target label is set as the "Many" class, some methods, particularly WaNet, exhibit improved performance when the target label is a "Few" class. This leads to WaNet performs better than our method in some cases with target label as ``Few'' class. One possible reason is that tail classes is dorminated by backdoored samples when target label is tail class for dirty-label attacks. Therefore, it becomes easier to learn the connection between target label and trigger.
% obtains a better ASR than other backdoor attacks using CUDA~\cite{ahn2022cuda}, especially on ''Many'' classes. %This observation demonstrates that our method can learn an effective backdoored data augmentation operations selector for head classes. The  
Note that clean-label attacks (LC~\cite{turner2019lc} and SIG~\cite{barni2019sig}) perform poorly when target labels belong to ``Few'' classes. The reason is that clean-label attacks only use samples from the target label to poison the training dataset. The number of samples from ``Few'' group is too small to achieve an effective backdoor attack. This observation is especially obvious on CIFAR100-LT because there are fewer samples of tail classes compared to CIFAR10-LT. The results show the impractical features of traditional clean-label attacks when target label is a tail class of an unbalanced dataset. 

Other results using Logits Adjustment (LA)~\cite{menon2020logitadj} can be found in Appendix. %The detailed results are presented in Appendix. Comparing results on Table~\ref{tab:cifar10-c} and Table~\ref{tab:la}, we observe that LA can improve ACC largely. This is consistent with the traditional long-tailed classification methods, which usually combine re-weight and re-sampling methods to achieve a better performance. However, LA sometimes can decrease attack performance (ASR). For example, the average ASR of IAB using LA is lower than ASR without using LA. A possible reason is that the decision boundary between backdoored samples and clean samples can be changed by LA.

\begin{table}[tp]
    \centering
    \medskip
\scalebox{0.7}{
    \begin{tabular}{@{\hspace{2mm}}c@{\hspace{2mm}}|
    @{\hspace{2mm}}c@{\hspace{2mm}}|
    @{\hspace{2mm}}c@{\hspace{2mm}}|
    @{\hspace{2mm}}c@{\hspace{2mm}}|
    @{\hspace{2mm}}c@{\hspace{2mm}}}
    \toprule
    Strength $q$ & All & Many & Medium & Few \\
    \midrule
0&94.70 / 72.67 &93.70 / 93.57 &95.53 / 71.33 &94.58 / 58.00 \\
1&96.99 / 74.30 &96.55 / 91.80 &97.50 / 73.67 &96.82 / 61.65 \\
2&98.18 / 76.68 &98.75 / 93.10 &98.40 / 76.23 &97.72 / 64.70 \\
3&94.27 / 75.38 &93.55 / 94.10 &95.13 / 74.60 &93.97 / 61.93 \\
4&92.77 / 76.04 &93.10 / 93.10 &95.57 / 73.20 &90.50 / 65.38 \\
    \bottomrule
    \end{tabular}
    }
    \vspace{-2mm}
    \caption{Analysis of data augmentation strength $q$.}
    \label{tab:ablation_q}
\end{table}

\begin{table}[tp]
    \centering
    \medskip
\scalebox{0.7}{
    \begin{tabular}{@{\hspace{2mm}}c@{\hspace{2mm}}|
    @{\hspace{2mm}}c@{\hspace{2mm}}|
    @{\hspace{2mm}}c@{\hspace{2mm}}|
    @{\hspace{2mm}}c@{\hspace{2mm}}|
    @{\hspace{2mm}}c@{\hspace{2mm}}}
    \toprule
$\alpha$&All&Many&Medium&Few\\
\midrule
0.01&39.49 / 74.50 &38.35 / 92.73 &35.00 / 73.53 &43.43 / 61.55 \\
0.05&79.29 / 76.08 &83.70 / 92.13 &80.07 / 73.50 &76.50 / 65.98 \\
0.1&98.18 / 76.68 &98.75 / 93.10 &98.40 / 76.23 &97.72 / 64.70 \\
0.15&98.51 / 73.98 &98.90 / 92.40 &98.67 / 73.20 &98.20 / 60.75 \\
0.2&99.09 / 74.48 &99.05 / 94.20 &99.43 / 72.17 &98.85 / 61.43 \\
    \bottomrule
    \end{tabular}
    }
\vspace{-2mm}
\caption{Analysis of trigger strength $\alpha$.}
\label{tab:ablation_alpha}
\end{table}

\subsection{Analysis}
We conduct ablation studies on CIFAR10-LT. The target label is set as 0. More defense experiments and analysis of other target labels can be found in Appendix. 

\textbf{Data Augmentation Strength $q$ in Backdoored Operation Selector.}
Data augmentation strength $q$ determines the operation set from which backdoored operation selector chooses weak data augmentation operations.
We fix other hyperparameters and change the values of $q$. The results are shown in Tab.~\ref{tab:ablation_q}. It is observed that ASRs first increase and then decrease with strength $q$ increasing. The model gets the best average ASR $98.2$ when $q$ is set as 2. Compared to the case without using data augmentation, ASR increases by $3.5$ percentages. After getting the best ASR, ASR will decrease quickly with data augmentation strength increases. 
Compared ``Many'' classes, using strong data augmentation operations can decrease ASR largely for ``Few'' classes.
% With a fixed data augmentation strength $q$, backdoored operations can choose $q$ data augmentation operations with strength $q$ according the probabilities predicted by network $h$. 
% Since network $h$ will predict high probabilities for weak data augmentation operations, 
Therefore, the results show that using several (less than 2) weak data augmentation operations can improve the trigger generator. However, using stronger data augmentation operations can hinder to learn an effective trigger generator especially for tail classes. This is consistent with one of our intuition.

\begin{table}[!t]
    \centering
    \medskip
\scalebox{0.7}{
    \begin{tabular}{@{\hspace{2mm}}c@{\hspace{2mm}}|
    @{\hspace{2mm}}c@{\hspace{2mm}}|
    @{\hspace{2mm}}c@{\hspace{2mm}}|
    @{\hspace{2mm}}c@{\hspace{2mm}}|
    @{\hspace{2mm}}c@{\hspace{2mm}}}
    \toprule
    $T$ & All & Many & Medium & Few \\
    \midrule
1&98.18 / 76.68 &98.75 / 93.10 &98.40 / 76.23 &97.72 / 64.70 \\
2&96.10 / 76.09 &94.25 / 92.83 &97.30 / 76.73 &96.12 / 63.05 \\
3&95.58 / 75.94 &94.95 / 93.17 &96.30 / 74.40 &95.35 / 64.18 \\
% 4&97.51 / 75.37 / 86.44&97.55 / 91.50 / 94.53&97.43 / 73.97 / 85.70&97.55 / 64.33 / 80.94\\
    \bottomrule
    \end{tabular}
    }
    \vspace{-2mm}
    \caption{Analysis of temperature $T$ in optimization objective $\mathcal{L}_{h}$ of backdoored selector.}
    \label{tab:ablation_T}
\end{table}

\begin{table}[tp]
    \centering
    \medskip
\scalebox{0.7}{
    \begin{tabular}{@{\hspace{2mm}}c@{\hspace{2mm}}|
    @{\hspace{2mm}}c@{\hspace{2mm}}|
    @{\hspace{2mm}}c@{\hspace{2mm}}|
    @{\hspace{2mm}}c@{\hspace{2mm}}|
    @{\hspace{2mm}}c@{\hspace{2mm}}}
    \toprule
    $\lambda_{div}$&All&Many&Medium&Few \\
    \midrule
% 0.0&96.89 / 75.43 / 86.16&97.40 / 92.37 / 94.89&96.93 / 76.17 / 86.55&96.60 / 62.18 / 79.39\\
0.01&98.18 / 76.68 &98.75 / 93.10 &98.40 / 76.23 &97.72 / 64.70 \\
0.05&96.33 / 74.39 &96.10 / 92.83 &97.23 / 72.63 &95.77 / 61.87 \\
0.1&94.82 / 75.41 &95.15 / 92.83 &94.57 / 75.83 &94.85 / 62.03 \\
0.5&91.04 / 71.49 &90.55 / 91.03 &92.43 / 67.33 &90.25 / 59.95 \\
    \bottomrule
    \end{tabular}
    }
\vspace{-2mm}
\caption{Analysis of diversity loss weight $\lambda_{div}$.}
\label{tab:ablation_div}
\end{table}

\textbf{Softmax Temperature $T$ in Optimization Objective of Network $h$.}
To analyze the effect of temperature $T$ on the optimization, we fix other hyperparameters and only change temperature $T$. The results are shown in Table~\ref{tab:ablation_T}. 
With temperature $T$ increasing, the average ASR decreases. Since increasing temperature $T$ can narrow the gap between probabilities output by network $h$, a possible reason is that choosing strong data augmentation operations is easier in the case of high temperature $T$. Therefore, selecting a low temperature $T$ can ensure backdoored operation selector to choose more approppriate data augmentation operations for training effective trigger generator.

% \begin{wrapfigure}{r}{0.65\textwidth}
\begin{figure}[!t]
\setlength\tabcolsep{0.5pt}%%
\centering
\footnotesize
\vspace{-4mm}
\begin{tabular}{cc}
    \includegraphics[width=.49\linewidth]{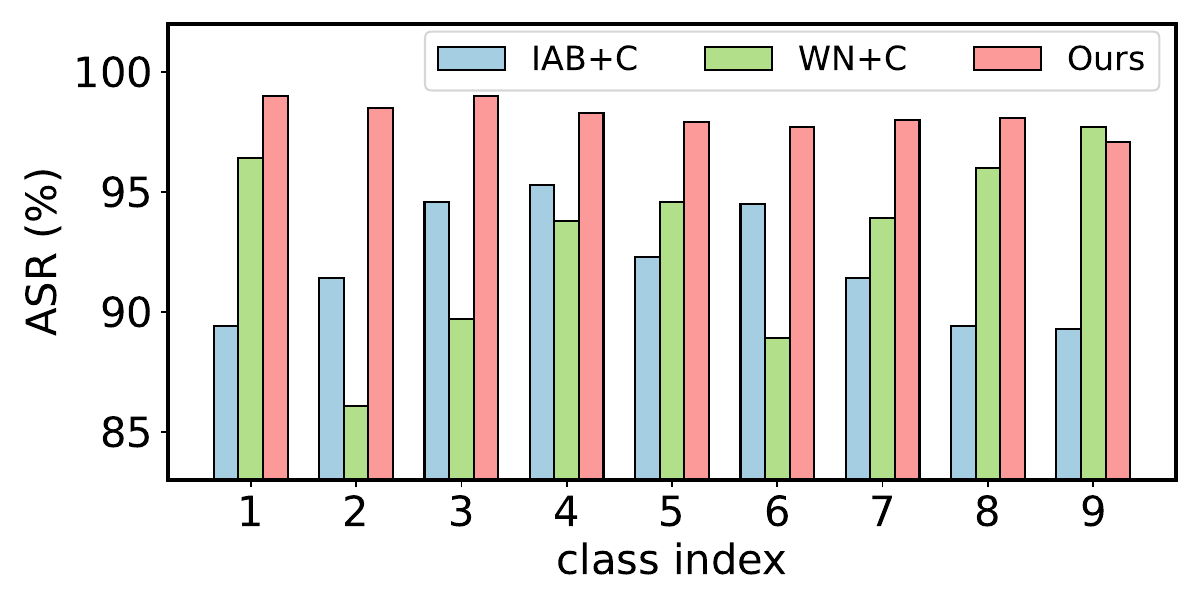} &
    \includegraphics[width=.49\linewidth]{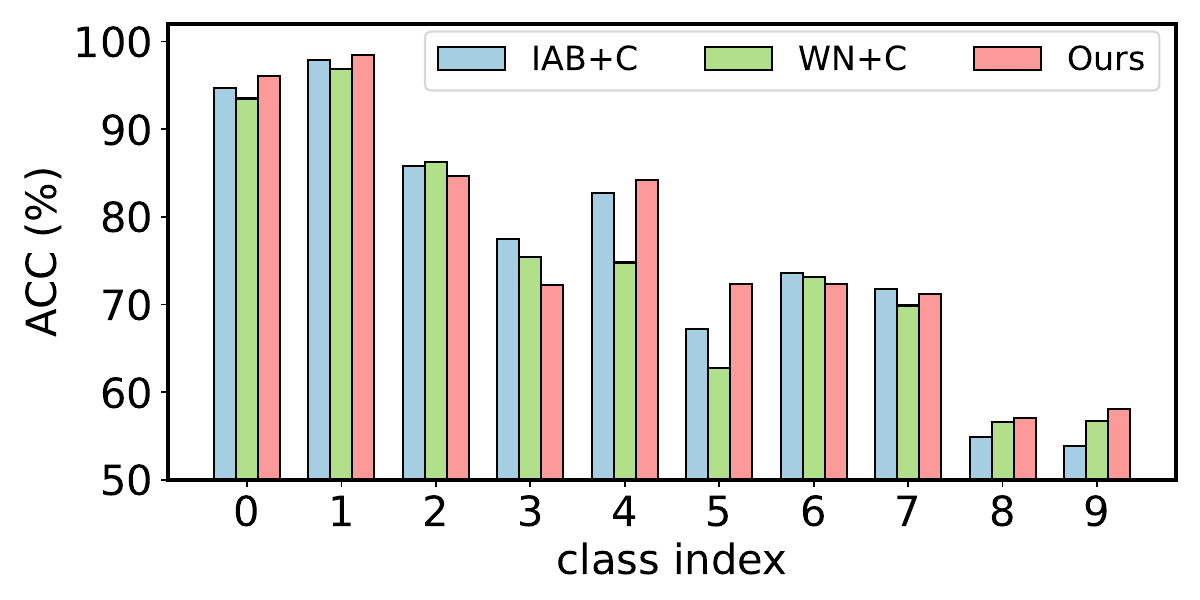} \\
    (a) {Class-wise ASR} &
    (b) {Class-wise ACC}
\end{tabular}
\vspace{-1mm}
    \caption{Class-wise performance comparison.}
    \label{fig:classwise}
\vspace{-4mm}
\end{figure}

\textbf{Weight ($\lambda_{div}$) of Trigger Diversity Loss.}
Trigger diversity loss ensures that trigger generator can generate sample-specific triggers instead of a constant trigger pattern. However, increasing trigger diversity makes trigger generator training more challenging. Due to the tradeoff between trigger diversity and trigger generator efficacy, we conduct an ablation study on the weight $\lambda_{div}$ of trigger diversity loss. We fix other hyperparameters and show the results in Table~\ref{tab:ablation_div}.
We can observe that increasing $\lambda_{div}$ causes the decrease of average ASR and ACC. The reason is that increasing weight of trigger diversity means the increasing the difficulty of training trigger generator. Therefore, the effectiveness (ASR) of generated triggers can be undermined. 
% This demonstrates that $\lambda_{div}$ can control the trade-off between effectiveness and diversity of triggers generated by trigger generator.

\textbf{Strength $\alpha$ of Trigger.}
We only tune the hyperparameter $\alpha$, and the results are shown in Table~\ref{tab:ablation_alpha}. The results show that increasing trigger strength can increase ASR of backdoored model. Also, increasing trigger strength can decrease average ACC of all classes slightly and decrease ACC largely for Few classes. A possible reason is that it is easier to flip images from other classes into target label. In other words, strong triggers can compress the feature space of tail classes.

\textbf{Class-wise Performance.}
We analyze the class-wise performance of three sample-specific backdoor attacks including IAB~\cite{nguyen2020IAB}, WaNet~\cite{nguyen2021wanet} and our method. Class-wise ASR and ACC are shown in Figure~\ref{fig:classwise}. The results show that our method achieve state-of-the-art ASR in the eight classes of all nine classes. For ACC, our method also perform better in most classes.
% especially tail classes. 
% More analysis and trigger visualizations can be found in supplementary.

% \paragraph{relation between DA and trigger generation}

\section{Conclusion}
\label{sec:conclusion}

In this paper, for the first time, we address the backdoor attack for long-tailed visual recognition. We propose a backdoor attack with Dynamic Data Augmentation Operations (D$^2$AO). The clean and backdoored selectors are designed to choose proper data augmentation operations for clean and backdoored samples separately. An auto-encoder trigger generator is used to generate stealthy trigger patterns. We conduct experiments on two long-tailed benchmarks and achieve state-of-the-art attack performance compared with other backdoor attacks.

\bibliography{aaai25}

\newpage
\setcounter{secnumdepth}{0}

\newpage
\appendix
\setcounter{secnumdepth}{1}

\begin{center}
\LARGE \textbf{Appendix}%\\[1em]
% \LARGE \textbf{Appendix - Long-Tailed Backdoor Attack Using Dynamic Data Augmentation Operations}%\\[1em]
\end{center}

\section{Experimental Details of Backdoor Attacks}
We compare our method with other four state-of-the-art stealthy backdoor attacks, including Label-consistent Backdoor Attack (LC), Sinusoidal Signal Backdoor Attack (SIG), Input-aware Backdoor Attack (IAB) and Warp-based Backdoor Attack (WN). LC and SIG are classic clean-label attacks. IAB and WaNet are representative sample-specific attacks. All experiments are conducted under the all-to-one setting. The implementation of our code is based on BackdoorBench~\cite{backdoorbench} V1~\footnote{\url{https://github.com/SCLBD/BackdoorBench/tree/v1}}. We follow the original paper (or code) of these attacks to conduct experiments, and the details of attack setting are introduced below:

\paragraph{Label-consistent Backdoor Attack (LC).} LC is a clean-label backdoor attack, and utilizes a trigger comprising four $3\times3$ checkerboards positioned at the four corners of an image. We follow the paper~\cite{turner2019lc} to generate adversarial perturbations using projected gradient descent (PGD)~\footnote{\url{https://github.com/MadryLab/label-consistent-backdoor-code}}. The adversarial model is trained with bounded in $l_{\rm inf}$ norm. For CIFAR10-LT, we follow the paper~\cite{turner2019lc} to set $\epsilon=16$. For CIFAR100-LT, we set $\epsilon=8$ to attack model successfully. We poison $50\%$ samples from target label to attack model for CIFAR10-LT. For CIFAR100-LT, we poison $80\%$ samples from target label to successfully attack models.

\paragraph{Sinusoidal Signal Backdoor Attack (SIG).} SIG is a clean-label attack with sinusoidal signal as triggers. In the case of CIFAR10-LT, we follow original paper~\cite{barni2019sig} to set $\Delta = 20$ and $f = 6$. For CIFAR100-LT, we set $\Delta = 40$ and $f = 6$ to successfully attack the model. Similar to LC, we poison $50\%$ samples from target label to attack model for CIFAR10-LT, while for CIFAR100-LT, we poison $80\%$ samples from target label to achieve successfully attacks.

\paragraph{Input-aware Backdoor Attack (IAB).} IAB is a sample-specific backdoor attack using two auto-encoders, one for generating triggers and the other for producing masks. 
% to generate triggers and masks respectively. 
Following the original code~\footnote{\url{https://github.com/VinAIResearch/input-aware-backdoor-attack-release}}, we first train a mask generator and then train the classification model and trigger generator simultaneously. The injection rate $\rho$ is set as $0.1$. We follow the paper to set $\rho_a$ as $0.1$ and $\rho_c$ as $0.1$ for both datasets.

\paragraph{Warp-based Backdoor Attack (WN).} WaNet employs elastic warping triggers to poison images. We follow the original code~\footnote{\url{https://github.com/VinAIResearch/Warping-based_Backdoor_Attack-release}} to train backdoored model with $\rho_a = 0.1$ and $\rho_n = 0.2$ for both CIFAR10-LT and CIFAR100-LT. The injection rate $\rho$ is also set as $0.1$.

\begin{table}[tp]
    \centering
    \scalebox{0.8}{
    \begin{tabular}{@{\hspace{2mm}}c@{\hspace{2mm}}|
    @{\hspace{2mm}}c@{\hspace{2mm}}|
    @{\hspace{2mm}}c@{\hspace{2mm}}|
    @{\hspace{2mm}}c@{\hspace{2mm}}|
    @{\hspace{2mm}}c@{\hspace{2mm}}|
    @{\hspace{2mm}}c@{\hspace{2mm}}}
    \toprule
    Detection & IAB & LC & SIG & WN & Ours \\
    \midrule

MNTD & 0.56 & 0.56 & 0.45 & 0.51 & 0.54\\
K-ARM & 0.50 & 0.50 & 0.35 & 0.45 & 0.45 \\
    \bottomrule
    \end{tabular}
    }
\caption{Backdoor detection accuracies on different backdoor attacked models.}
\label{tab:detection}
\end{table}

\begin{figure}[tp]
\centering
\begin{subfigure}[t]{0.49\linewidth}
\centering
    \includegraphics[width=\linewidth]{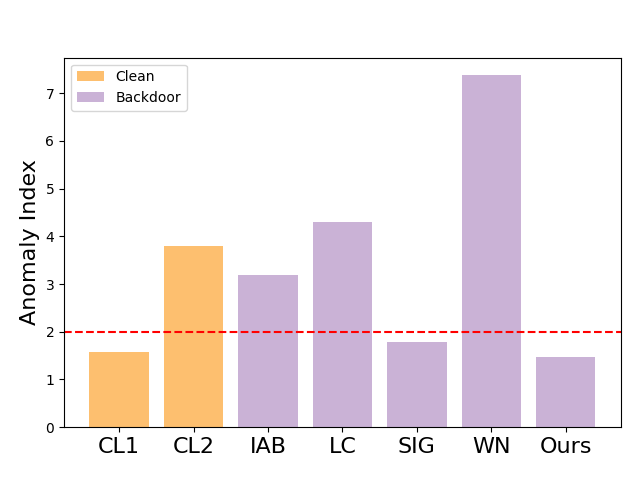}
    \caption{Neural Cleanse (NC)~\cite{wang2019neural}.}
    \label{fig:nc}
\end{subfigure}
\begin{subfigure}[t]{0.49\linewidth}
    \centering
    \includegraphics[width=\linewidth]{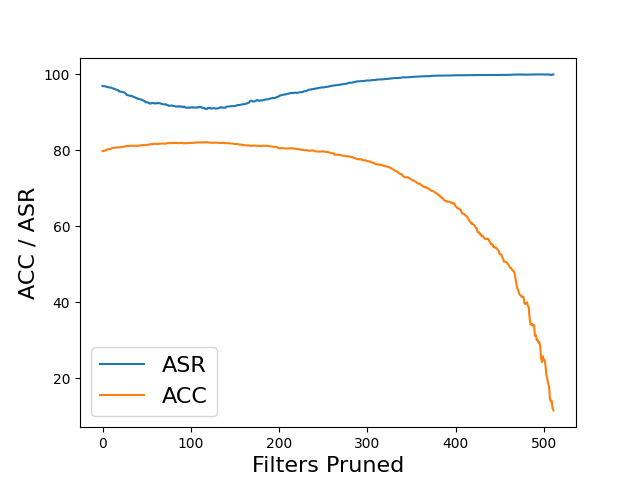}
    \caption{Fine-Pruning (FP)~\cite{liu2018fine}.}
    \label{fig:fp}
\end{subfigure}
    \caption{Results of resilience against Neural Cleanse and Fine-Pruning. (a) Models are flagged as backdoorded if Anomaly Index exceeds 2. (b) Our attack is resilient against fine-pruning.}
\end{figure}

% \begin{figure}[tp]
% \centering
% \begin{subfigure}[t]{0.49\linewidth}
% \centering
%     \includegraphics[width=\linewidth]{fig/nc_head.png}
%     \caption{Neural Cleanse (NC).}
%     \label{fig:nc}
% \end{subfigure}
% \begin{subfigure}[t]{0.49\linewidth}
%     \centering
%     \includegraphics[width=\linewidth]{fig/fp_0.95.png}
%     \caption{Fine-Pruning (FP).}
%     \label{fig:fp}
% \end{subfigure}
%     \caption{Results of resilience against Neural Cleanse and Fine-Pruning. (a) Models are flagged as backdoorded if Anomaly Index exceeds 2. (b) Our attack is resilient against fine-pruning.}
% \end{figure}

\begin{table*}[tp!]  
    \centering
    \scalebox{0.8}{
    \begin{tabular}{c|c|c|c|c|c|c|c|c|c|c}
    \toprule
    \multirow{2}{*}{Attack} & \multicolumn{2}{c|}{IAB} & \multicolumn{2}{c|}{LC} & \multicolumn{2}{c|}{SIG} & \multicolumn{2}{c|}{WN} & \multicolumn{2}{c}{Ours} \\
    \cmidrule{2-11}
& before & after & before & after & before & after & before & after & before & after \\
\midrule
ACC & 76.00 & 57.09 & 78.88 & 73.58 & 75.37 & 67.04 & 74.60 & 53.64 & 76.68 & 73.87\\
ASR &  91.95 & 80.50 & 85.43 & 15.80 & 93.15 & 94.00 & 93.01 & 67.56 & 98.17 & 96.37\\
    \bottomrule
    \end{tabular}
    }
\caption{Backdoor mitigation results using CLP~\cite{zheng2022data} on different backdoor attack models.}
\label{tab:clp}
\end{table*}

\section{Experiments of Resilience Against Backdoor Defenses}

% Backdoor defense can be divided into three categories: backdoored model detection, backdoored sample detection and backdoor mitigation. Backdoored model detection aims to determine whether a suspicious model is a backdoored model. 
Backdoor defense can be categorized into two main types: backdoor detection and backdoor mitigation. Backdoor detection focuses on identifying whether a given model is a backdoored model.
% Backdoored sample detection focuses on determining whether an input sample is backdoored input. 
On the other hand, backdoor mitigation aims to reduce the impact of backdoor attacks, typically by decreasing the Attack Success Rate (ASRs) while maintaining clean accuracies (ACCs) at an acceptable level. To assess the resilience against these defense approaches, we select representative methods from both categories and evaluate defense performance against five backdoor attacks, including our proposed method, on the CIFAR10-LT dataset. For a fair comparison, we integrate the other four attacks with the state-of-the-art data augmentation based long-tailed method called CUDA~\cite{ahn2022cuda}.

\begin{table*}[tp!]
\footnotesize
% \small
    \centering
    % \resizebox{\linewidth}{!}{
    \scalebox{0.9}{
    \renewcommand{\tabcolsep}{0.1cm}
    \begin{tabular}{c|c|c|c|c|c|c|c|c|c|c|c|c|c|c}
\toprule
\multirow{2}{*}[-0.4em]{Metric} 
% & \multirow{2}{*}{\begin{tabular}[c]{@{}c@{}}\textbf{Backdoor}\\ \textbf{Attacks}\end{tabular}} 
& \multirow{2}{*}[-0.4em]{Attack}
& \multicolumn{4}{c|}{Target Label: Many} & \multicolumn{4}{c|}{Target Label: Medium} & \multicolumn{4}{c|}{Target Label: Few} 
& \multirow{2}{*}[-0.4em]{Avg}\\ 
\cmidrule{3-14} 
& & Many & Med. & Few & All 
& Many & Med. & Few & All 
& Many & Med. & Few & All 
\\ 
\midrule
\multirow{5}{*}[-0.3em]{ACC} 
% \textbf{BN+C+LA}& 90.33 & 76.78 & 80.00 & 82.19 & 90.22 & 76.11 & 80.42 & 82.05 & 89.92 & 77.17 & 78.06 & 81.36 &81.81\\
% \textbf{BN+LA}& 90.00 & 71.00 & 69.00 & 75.93 & 89.00 & 70.33 & 68.42 & 75.14 & 89.33 & 70.58 & 67.06 & 74.81 &75.25\\
% \textbf{Bl+C+LA}& 91.89 & 78.56 & 81.67 & 83.72 & 90.44 & 78.00 & 82.08 & 83.36 & 88.17 & 76.33 & 82.06 & 82.15 &82.98\\
% \textbf{Bl+LA}& 91.22 & 75.33 & 72.58 & 79.05 & 91.44 & 75.33 & 72.42 & 79.02 & 91.33 & 74.92 & 72.31 & 78.75 &78.92\\
& IAB+LA & 88.88 & 76.74 & 83.53 & 83.10 & 89.17 & 75.63 & 82.87 & 82.59 & 88.91 & 76.24 & 82.67 & 82.61 &82.75\\
% \textbf{IAB+LA}& 91.36 & 75.51 & 71.18 & 78.53 & 91.09 & 75.09 & 70.92 & 78.22 & 90.85 & 74.76 & 71.36 & 78.23 &78.32\\
& LC+LA & 89.67 & 78.78 & 83.83 & 84.13 & 91.44 & 76.33 & 84.00 & 83.96 & 92.08 & 78.75 & 82.56 & 84.21 &84.11\\
% \textbf{LC+LA}& 90.11 & 76.33 & 73.58 & 79.46 & 92.33 & 73.22 & 73.67 & 79.19 & 91.92 & 75.67 & 70.81 & 78.60 &79.03\\
& SIG+LA & 89.33 & 79.11 & 83.67 & 84.01 & 91.78 & 75.00 & 82.67 & 83.13 & 91.58 & 78.42 & 80.75 & 83.30 &83.46\\
% \textbf{SIG+LA}& 90.33 & 75.33 & 74.75 & 79.62 & 92.33 & 72.11 & 74.42 & 79.10 & 92.08 & 74.83 & 70.81 & 78.44 &78.99\\
& WN+LA & 89.66 & 74.97 & 79.38 & 81.14 & 89.51 & 76.28 & 78.54 & 81.15 & 89.56 & 73.93 & 78.06 & 80.27 &80.79\\
% \textbf{WN+LA}& 89.78 & 69.54 & 67.46 & 74.78 & 90.09 & 69.71 & 67.80 & 75.06 & 89.43 & 68.88 & 67.95 & 74.67 &74.82\\
\cmidrule{2-15}
& Ours+LA & 90.72 & 77.19 & 80.96 & 82.76 & 89.46 & 77.00 & 81.70 & 82.62 & 88.44 & 75.44 & 82.82 & 82.29 &82.53\\

\midrule
\midrule
\multirow{5}{*}[-0.3em]{ASR} 
% \textbf{BN+C+LA}& 100.00 & 100.00 & 96.17 & 98.26 & 100.00 & 99.00 & 97.00 & 98.42 & 100.00 & 100.00 & 98.33 & 99.38 &98.75\\
% \textbf{BN+LA}& 97.33 & 91.11 & 80.33 & 87.66 & 95.44 & 88.17 & 79.75 & 86.78 & 96.92 & 90.92 & 84.50 & 90.71 &88.61\\
% \textbf{Bl+C+LA}& 99.00 & 98.22 & 91.08 & 95.19 & 99.56 & 99.17 & 94.08 & 96.95 & 99.83 & 99.33 & 95.92 & 98.30 &96.96\\
% \textbf{Bl+LA}& 99.17 & 98.56 & 93.58 & 96.54 & 99.67 & 99.50 & 96.08 & 97.99 & 99.92 & 99.75 & 97.58 & 98.99 &97.96\\
& IAB+LA & 89.15 & 79.64 & 64.06 & 74.83 & 97.64 & 93.78 & 84.71 & 91.04 & \textbf{99.57} & 98.44 & 94.50 & 97.50 &88.76\\
% \textbf{IAB+LA}& 93.68 & 86.32 & 78.49 & 84.48 & 99.00 & 95.37 & 92.14 & 95.14 & 99.66 & 98.68 & 96.53 & 98.29 &93.20\\
& LC+LA & 88.67 & 88.33 & 82.25 & 85.74 & 90.11 & \textbf{99.33} & 93.25 & 93.59 & 29.08 & 42.08 & 22.33 & 31.17 &66.27\\
% \textbf{LC+LA}& 94.67 & 91.11 & 93.75 & 93.05 & 84.89 & 95.33 & 88.42 & 88.74 & 22.00 & 34.25 & 22.33 & 26.21 &65.02\\
& SIG+LA & 94.67 & 87.78 & 90.25 & 90.47 & 93.56 & 93.17 & 92.17 & 92.82 & 81.00 & 68.75 & 68.83 & 72.84 &84.12\\
% \textbf{SIG+LA}& 85.67 & 78.67 & 82.67 & 82.00 & 85.00 & 91.67 & 84.17 & 86.06 & 58.50 & 47.50 & 45.83 & 50.58 &70.65\\
& WN+LA & 93.88 & 94.38 & 89.62 & 92.16 & 96.59 & 96.70 & \textbf{95.67} & 96.21 & 97.75 & \textbf{98.83} & \textbf{97.67} & \textbf{98.09} &95.74\\
% \textbf{WN+LA}& 93.58 & 95.74 & 94.12 & 94.54 & 97.23 & 98.40 & 97.38 & 97.56 & 98.02 & 98.67 & 98.08 & 98.26 &96.93\\
\cmidrule{2-15}
& Ours+LA & \textbf{97.63} & \textbf{95.99} & \textbf{95.67} & \textbf{96.21} & \textbf{98.36} & 95.58 & 94.91 & \textbf{96.21} & 98.72 & 96.62 & 93.48 & 96.28 &\textbf{96.24}\\
\bottomrule
\end{tabular}
    }
\caption{Comparison results using Logits Adjustment (LA) on CIFAR10-LT.}
\label{tab:cifar10-la}
\end{table*}

\begin{table*}[tp!]
\footnotesize
    \centering
    \scalebox{0.9}{
    \renewcommand{\tabcolsep}{0.1cm}
    \begin{tabular}{c|c|c|c|c|c|c|c|c|c|c|c|c|c|c}
\toprule
\multirow{2}{*}[-0.4em]{Metric} 
% & \multirow{2}{*}{\begin{tabular}[c]{@{}c@{}}\textbf{Backdoor}\\ \textbf{Attacks}\end{tabular}} 
& \multirow{2}{*}[-0.4em]{Attack}
& \multicolumn{4}{c|}{Target Label: Many} & \multicolumn{4}{c|}{Target Label: Medium} & \multicolumn{4}{c|}{Target Label: Few} 
& \multirow{2}{*}[-0.4em]{Avg}\\ 
\cmidrule{3-14} 
& & Many & Med. & Few & All 
& Many & Med. & Few & All 
& Many & Med. & Few & All 
\\ 
\midrule
\multirow{5}{*}{ACC} 
% \textbf{BN+C+LA}& 65.22 & 57.74 & 50.13 & 57.62 & 66.39 & 57.84 & 49.94 & 57.98 & 65.40 & 57.39 & 49.34 & 57.30 &57.63\\
% \textbf{BN+LA}& 59.90 & 51.07 & 42.63 & 51.12 & 61.16 & 51.54 & 42.49 & 51.64 & 61.66 & 52.26 & 41.91 & 51.84 &51.53\\
% \textbf{Bl+C+LA}& 67.33 & 59.31 & 51.78 & 59.40 & 67.41 & 58.26 & 52.19 & 59.22 & 68.02 & 59.27 & 51.27 & 59.44 &59.35\\
% \textbf{Bl+LA}& 62.52 & 53.10 & 43.40 & 52.91 & 63.45 & 53.18 & 44.62 & 53.66 & 63.66 & 53.66 & 44.04 & 53.69 &53.42\\
& IAB+LA & 65.62 & 58.33 & 51.26 & 58.33 & 65.71 & 58.38 & 51.03 & 58.30 & 65.51 & 58.47 & 50.54 & 58.10 &58.24\\
% \textbf{IAB+LA}& 63.92 & 53.99 & 45.38 & 54.34 & 63.69 & 54.20 & 44.77 & 54.13 & 63.75 & 53.81 & 44.97 & 54.08 &54.18\\
& LC+LA & 67.26 & 60.16 & 53.15 & 60.12 & 68.02 & 60.16 & 52.55 & 60.17 & 68.48 & 60.49 & 53.15 & 60.63 &60.31\\
% \textbf{LC+LA}& 64.18 & 55.10 & 45.51 & 54.84 & 64.81 & 54.41 & 45.95 & 54.97 & 63.91 & 54.57 & 44.93 & 54.37 &54.72\\
& SIG+LA & 67.49 & 59.84 & 52.74 & 59.95 & 66.07 & 56.92 & 51.40 & 58.06 & 67.71 & 59.65 & 51.24 & 59.45 &59.16\\
% \textbf{SIG+LA}& 63.67 & 54.86 & 45.34 & 54.53 & 63.56 & 53.88 & 45.29 & 54.15 & 62.12 & 53.51 & 43.53 & 52.96 &53.87\\
& WN+LA & 68.02 & 60.22 & 52.65 & 60.22 & 67.35 & 59.76 & 53.40 & 60.10 & 67.67 & 59.43 & 52.56 & 59.81 &60.04\\
% \textbf{WN+LA}& 65.02 & 55.24 & 46.12 & 55.36 & 64.96 & 55.55 & 46.27 & 55.50 & 63.95 & 54.86 & 45.79 & 54.78 &55.21\\
\cmidrule{2-15}
& Ours+LA & 66.22 & 57.89 & 50.57 & 58.15 & 66.14 & 58.91 & 49.65 & 58.15 & 65.97 & 59.25 & 50.00 & 58.32 &58.21\\
\midrule
\midrule
\multirow{5}{*}{ASR} 
% \textbf{BN+C+LA}& 99.96 & 99.21 & 99.60 & 99.59 & 99.88 & 98.81 & 99.25 & 99.32 & 99.99 & 99.52 & 99.79 & 99.76 &99.56\\
% \textbf{BN+LA}& 98.20 & 96.39 & 97.73 & 97.43 & 97.95 & 96.26 & 97.20 & 97.14 & 98.16 & 96.52 & 98.04 & 97.57 &97.38\\
% \textbf{Bl+C+LA}& 99.55 & 99.34 & 99.48 & 99.46 & 99.67 & 99.30 & 99.62 & 99.53 & 99.43 & 99.34 & 99.51 & 99.43 &99.47\\
% \textbf{Bl+LA}& 100.00 & 99.89 & 100.00 & 99.96 & 100.00 & 99.88 & 99.98 & 99.95 & 99.97 & 99.83 & 99.93 & 99.91 &99.94\\
& IAB+LA & 97.95 & 96.96 & 96.46 & 97.11 & \textbf{98.92} & 97.92 & 97.80 & 98.21 & \textbf{99.05} & \textbf{98.23} & 98.04 & 98.44 &97.93\\
% \textbf{IAB+LA}& 98.79 & 98.59 & 98.12 & 98.49 & 99.34 & 99.15 & 98.98 & 99.15 & 99.52 & 99.26 & 99.19 & 99.32 &98.99\\
& LC+LA & 42.30 & 41.31 & 46.70 & 43.48 & 61.91 & 59.38 & 64.09 & 61.84 & 1.12 & 1.74 & 1.71 & 1.52 &35.27\\
% \textbf{LC+LA}& 44.32 & 43.15 & 49.24 & 45.62 & 59.35 & 59.59 & 62.76 & 60.60 & 11.51 & 14.74 & 14.16 & 13.47 &39.63\\
& SIG+LA & 94.41 & 94.33 & 96.00 & 94.93 & 93.58 & 93.28 & 95.42 & 94.11 & 91.02 & 92.03 & 93.13 & 92.06 &93.69\\
% \textbf{SIG+LA}& 90.20 & 89.58 & 90.67 & 90.15 & 84.98 & 84.28 & 86.17 & 85.16 & 86.93 & 88.44 & 87.88 & 87.75 &87.69\\
& WN+LA & 92.63 & 93.86 & 90.50 & 92.31 & 93.09 & 94.34 & 90.93 & 92.75 & 94.91 & 96.25 & 92.69 & 94.62 &93.24\\
% \textbf{WN+LA}& 96.51 & 96.98 & 95.39 & 96.28 & 97.07 & 97.81 & 95.95 & 96.93 & 98.29 & 98.70 & 97.31 & 98.10 &97.11\\
\cmidrule{2-15}
& Ours+LA & \textbf{98.80} & \textbf{98.48} & \textbf{98.45} & \textbf{98.57} & 98.39 & \textbf{98.27} & \textbf{98.38} & \textbf{98.35} & 98.97 & 98.15 & \textbf{98.48} & \textbf{98.54} &\textbf{98.49}\\
\bottomrule
\end{tabular}
    }
    \caption{Comparison results using Logits Adjustment (LA) on CIFAR100-LT.}
\label{tab:cifar100-la}
\end{table*}

\paragraph{Resilience Against Backdoor Detection.}
Neural Cleanse (NC)~\cite{wang2019neural}, Meta
Neural Trojan Detection (MNTD)~\cite{xu2021meta} and K-ARM~\cite{shen2021karm} represent three classic methods for backdoor detection. 

Neural Cleanse (NC)~\cite{wang2019neural} optimizes a trigger pattern for each label to convert all clean samples into this specified label. It then employs an outlier detection algorithm to identify triggers smaller than others. NC uses the Anomaly Index for evaluation. If the Anomaly Index exceeds 2, the model is flagged as backdoored. The results of five backdoor attacks' resiliences against NC are shown in Fig.~\ref{fig:nc}. All backdoor attacks are integrated with the state-of-the-art data augmentation based long-tailed method CUDA~\cite{ahn2022cuda}. The target label is set as 0 (head class). We train two types of clean models: ``CL1'' denotes training without data augmentation, while ``CL2'' denotes training with CUDA. We can observe that using CUDA increases Anomaly Index from the clean model. Therefore, most other backdoor attacks exhibiting Anomaly Index values exceeding 2 due to the integration with CUDA. In contrast, our method utilizes dynamic data augmentation operations instead of CUDA, allowing it to evade detection by NC.

Meta Neural Trojan Detection (MNTD)~\cite{xu2021meta} trains a mete-classifier to discern whether a model is backdoored or clean. The meta-classifier is trained via a jumbo-learning strategy which samples a set of backdoored model following a general distribution.
K-ARM~\cite{shen2021karm} iteratively and stochastically selects the most promising target labels to optimize the trigger. Both methods evaluate results based on detection accuracy. For each backdoor attack, we train 10 backdoored models with 10 class labels as target labels separately, along with 10 clean models. MNTD is conducted 3 times to compute the average accuracies. The results, presented in Tab.~\ref{tab:detection}, indicate that neither detection method effectively distinguishes backdoored and clean models. This suggests that all backdoor attacks can evade detection by MNTD and K-ARM, likely due to the inapplicability of these detection methods to backdoored models trained on long-tailed datasets.

\paragraph{Resilience Against Backdoor Mitigation.}
We choose a classic backdoor mitigation method called Fine-Pruning (FP)~\cite{liu2018fine} and a recent method called CLP~\cite{zheng2022data}.

Fine-Pruning (FP)~\cite{liu2018fine} prunes neurons with low activations for clean samples. These neurons typically correspond to backdoor behaviors. We train a backdoored model using our proposed attack with target label as 0 (a head class). The Fig.~\ref{fig:fp} shows the ASRs and ACCs with respect to the number of pruned filters. With increased number of pruned filters, ASR of our attack will not decrease and ACC will decrease. 
Therefore, our method is resilient against FP. 
%With the same filters pruned, our attack method generally retains higher ASR than all other attacks.
%However, our method exhibits a slower decrease compared to the other four backdoor attacks. Therefore, our method demonstrates superior resilience against FP.

CLP~\cite{zheng2022data} detects potential backdoor channel through Channel Lipschitz Constant (CLC), and then prune detected backdoor channel. We follow the original paper to set hyperparameter $u$ as 3 for CIFAR10-LT dataset. We present mitigation results using CLP on five backdoor attacks on CIFAR10-LT in Table~\ref{tab:clp}. From the results, we can observe that CLP usually reduces ACC and ASR simultaneously. Therefore, most backdoor attacks is resilient against CLP under the long-tailed visual recognition. One possible reason is that long-tailed backdoor models do not follow the channel Lipschitz condition. This encourages to explore backdoor defenses on long-tailed dataset.

\section{Comparison Results Using Logits Adjustments (LA)}
\label{sec:la}
We present comparison results on CIFAR10-LT and CIFAR100-LT using Logits Adjustments (LA) in Table~\ref{tab:cifar10-la} and Table~\ref{tab:cifar100-la}, respectively. Comparing to the results without using LA in the main paper, we observe that LA can increase ACC largely. This is consistent with the traditional long-tailed visual recognition methods, which usually integrate re-sampling and re-weighting methods to improve the classification accuracy. However, LA sometimes hinders the attack performance (ASR). For example, the average attack performance of IAB using LA decreases. A possible reason is that LA changes the decision boundary between clean images and backdoored images.

\begin{figure}[tp]
    \centering
    \includegraphics[width=0.8\columnwidth]{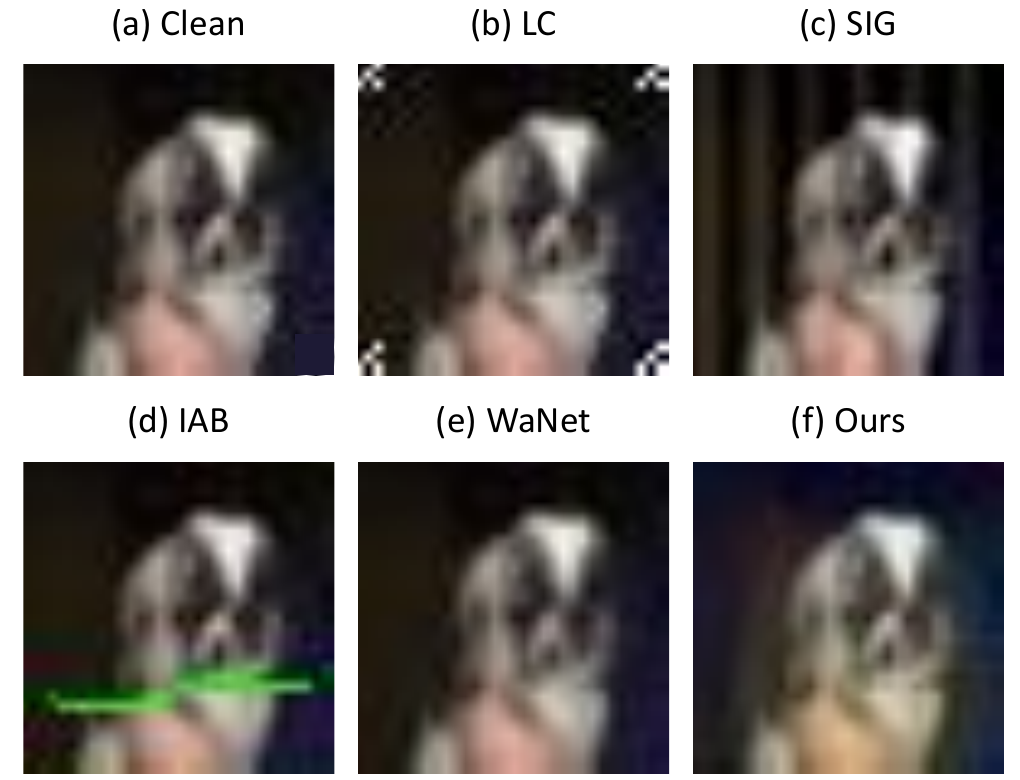}
    \caption{Backdoored images using different attacks.}
    \label{fig:triggers}
\end{figure}

\section{More Experimental Analysis}
%\subsection{Results on ImageNet-LT}
%\subsection{Discussion: trigger generation vs.~augmentation.}
% \cc{If you can, discuss a bit about how trigger and augmentation mutually influence. This could also be put at the end of the experiment section.}

%\subsection{Resilience against existing defense and Adaptive Defense}

\subsection{Visualization of backdoored images }
We visualize the backdoored images generated by different backdoor attacks in Fig.~\ref{fig:triggers}. It's evident that the triggers used by two clean backdoor attacks, LC and SIG, lack stealthiness. The stealthiness of these two attacks is that they do not change ground-truth labels. While IAB is a sample-specific backdoor attack, its trigger is still discernible to human inspectors. In contrast, the triggers generated by WaNet and our method exhibit a higher level of stealthiness. The difference is that WaNet subtly alters the shape of objects (the dog is thiner than that in the original image), while our method introduces slight color changes..

%\subsection{Visualize chosen policies}

\begin{table}[tp]
    \centering
    \scalebox{0.8}{
    \begin{tabular}{@{\hspace{2mm}}c@{\hspace{2mm}}|
    @{\hspace{2mm}}c@{\hspace{2mm}}|
    @{\hspace{2mm}}c@{\hspace{2mm}}|
    @{\hspace{2mm}}c@{\hspace{2mm}}|
    @{\hspace{2mm}}c@{\hspace{2mm}}}
    \toprule
    Strength $q$ & All & Many & Medium & Few \\
    \midrule

0&95.49 / 76.23 &95.80 / 92.03 &97.00 / 75.97 &94.50 / 64.58 \\
1&93.51 / 77.03 &93.27 / 93.23 &95.35 / 76.13 &92.77 / 65.55 \\
2&97.26 / 75.90 &97.43 / 93.07 &97.40 / 74.93 &97.05 / 63.75 \\
3&94.82 / 75.97 &95.03 / 92.80 &95.20 / 75.87 &94.47 / 63.42 \\
4&97.64 / 74.69 &98.10 / 92.33 &97.80 / 75.37 &97.23 / 60.95 \\
    \bottomrule
    \end{tabular}
    }
\caption{Analysis of data augmentation strength $q$ in backdoored operation selector.}
\label{tab:ablation_q_medium}

\end{table}

\begin{table}[tp]
    \centering
    \scalebox{0.8}{
    \begin{tabular}{@{\hspace{2mm}}c@{\hspace{2mm}}|
    @{\hspace{2mm}}c@{\hspace{2mm}}|
    @{\hspace{2mm}}c@{\hspace{2mm}}|
    @{\hspace{2mm}}c@{\hspace{2mm}}|
    @{\hspace{2mm}}c@{\hspace{2mm}}}
    \toprule
    $T$ & All & Many & Medium & Few \\
    \midrule
2&97.82 / 74.86 &98.07 / 92.97 &97.80 / 75.07 &97.65 / 61.13 \\
1&97.26 / 75.90 &97.43 / 93.07 &97.40 / 74.93 &97.05 / 63.75 \\
3&97.89 / 75.77 &98.47 / 92.33 &98.20 / 72.87 &97.30 / 65.52 \\
% 4&95.78 / 75.07 / 85.42&95.87 / 93.80 / 94.84&95.85 / 73.57 / 84.71&95.67 / 62.15 / 78.91\\

    \bottomrule
    \end{tabular}
    }
\caption{Results on CIFAR10-LT when changing temperature $T$ in optimization objective $\mathcal{L}_{h}$ of backdoored operation selector.}
    \label{tab:ablation_T_medium} 
\end{table}

\subsection{Analysis on other target labels}
In our main paper, we conduct analysis experiments on CIFAR10-LT with target label as 0 (``Many'' classes). we extend our analysis with the target label set as 4 (``Medium'' classes).. These experiments follow a similar structure to those presented in the main paper. Below are the details of our ablation studies:

\paragraph{Data Augmentation Strength $q$ in Backdoored Operation Selector.}
Similar to the experiments conducted in the main paper, we keep the other hyperparameters fixed. The results are presented in Tab.~\ref{tab:ablation_q_medium}. 
We can observe that the model gets higher ASRs when $q$ is set as $2$ or $4$. Compared to models without using data augmentation ($q = 0$), ASRs increase about $2\%$. This observation underscores the effectiveness of employing weak data augmentation policies, consistent with our findings in the main paper.
Since ASRs can be higher for both $q=2$ and $q=4$, data augmentation strengths can be different for different target labels. Tuning hyperparameter $q$ for different target labels can get better overall performance.

\paragraph{Softmax Temperature $T$ in Optimization Objective of Network $h$.}
We fix other hyperparameters and only change temperature $T$. The results are shown in Tab.~\ref{tab:ablation_T_medium}. 
From the results, we can see that changing temperature $T$ does not significantly affect average ASRs. This suggests that the choice of $T$ does not play a crucial role in achieving better performance on ASRs when the target label is 4.

\begin{table}[tp]
    \centering
    \scalebox{0.8}{
    \begin{tabular}{@{\hspace{2mm}}c@{\hspace{2mm}}|
    @{\hspace{2mm}}c@{\hspace{2mm}}|
    @{\hspace{2mm}}c@{\hspace{2mm}}|
    @{\hspace{2mm}}c@{\hspace{2mm}}|
    @{\hspace{2mm}}c@{\hspace{2mm}}}
    \toprule
    $\lambda_{div}$&All&Many&Medium&Few \\
    \midrule
0.01&97.26 / 75.90 &97.43 / 93.07 &97.40 / 74.93 &97.05 / 63.75 \\
0.05&97.39 / 75.86 &97.70 / 92.80 &98.70 / 75.73 &96.50 / 63.25 \\
% 0.0&97.81 / 75.85 / 86.83&98.90 / 93.53 / 96.22&97.55 / 74.83 / 86.19&97.12 / 63.35 / 80.23\\
0.1&97.59 / 74.37 &97.27 / 91.73 &98.15 / 75.97 &97.55 / 60.15 \\
0.5&89.02 / 76.06 &90.30 / 93.73 &91.00 / 74.30 &87.08 / 64.12 \\
    \bottomrule
    \end{tabular}
    }
\caption{Analysis of trigger diversity loss weight $\lambda_{div}$.}
\label{tab:ablation_div_medium}
\end{table}

\begin{table}[tp!]
    \centering
    \scalebox{0.8}{
    \begin{tabular}{@{\hspace{2mm}}c@{\hspace{2mm}}|
    @{\hspace{2mm}}c@{\hspace{2mm}}|
    @{\hspace{2mm}}c@{\hspace{2mm}}|
    @{\hspace{2mm}}c@{\hspace{2mm}}|
    @{\hspace{2mm}}c@{\hspace{2mm}}}
    \toprule
$\alpha$&All&Many&Medium&Few\\
\midrule
0.01&19.18 / 74.13 &12.10 / 92.80 &25.80 / 74.33 &21.18 / 59.98 \\
0.05&82.28 / 76.38 &82.73 / 94.07 &85.00 / 75.37 &80.58 / 63.87 \\
0.1&97.26 / 75.90 &97.43 / 93.07 &97.40 / 74.93 &97.05 / 63.75 \\
0.15&98.89 / 75.66 &98.90 / 93.77 &99.20 / 73.47 &98.72 / 63.73 \\
0.2&98.14 / 75.98 &98.77 / 93.17 &98.25 / 74.43 &97.62 / 64.25 \\
    \bottomrule
    \end{tabular}
    }
\caption{Analysis of trigger strength $\alpha$.}
\label{tab:ablation_alpha_medium}
\end{table}

\begin{figure}[tp!]
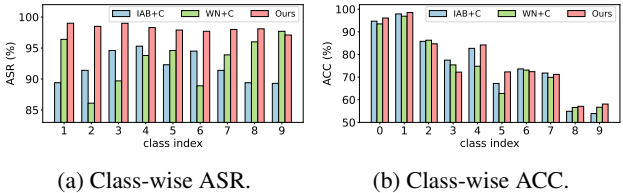

\centering
\begin{subfigure}[t]{0.49\linewidth}
\centering
    \includegraphics[width=\linewidth]{fig/per-class-asr.pdf}
    \caption{Class-wise ASR.}
    \label{fig:classwise-asr-medium}
\end{subfigure}
\begin{subfigure}[t]{0.49\linewidth}
    \centering
    \includegraphics[width=\linewidth]{fig/per-class-acc.pdf}
    \caption{Class-wise ACC.}
    \label{fig:classwise-acc-medium}
\end{subfigure}
    \caption{Class-wise performance compared to other two attacks when target label is 4.}
\end{figure}

\paragraph{Weight ($\lambda_{div}$) of Trigger Diversity Loss.}
The results of varying $\lambda_{div}$ are presented in Tab.~\ref{tab:ablation_div_medium}.
It is noticeable that ASR experiences a significant decrease with a larger value of $\lambda_{div}$. The reason is that increasing $\lambda_{div}$ can increase the difficulty of training trigger generator. Essentially, $\lambda_{div}$ represents the trade-off between the effectiveness and diversity of the trigger generator, which aligns with the observations made in the main paper.

\paragraph{Strength $\alpha$ of Trigger.}
We conducted experiments with fixed hyperparameters while adjusting $\alpha$, and the results are summarized in Tab.~\ref{tab:ablation_alpha_medium}. It is observed that ASRs will increase when increasing trigger strength $\alpha$. This observation can be attributed to the heightened visibility of the trigger as its strength increases. A stronger trigger is more conspicuous, making it easier for the classifier to discern and learn its features, thus resulting in higher ASRs.

\paragraph{Class-wise Performance.}
We further analyzed the class-wise performance, comparing our method with two other sample-specific backdoor attacks, namely IAB~\cite{nguyen2020IAB} and WaNet~\cite{nguyen2021wanet}. Class-wise ASR and ACC are depicted in Figure~\ref{fig:classwise-asr-medium} and Fig.~\ref{fig:classwise-acc-medium}, respectively. From Fig.~\ref{fig:classwise-asr-medium}, it is evident that our method achieves state-of-the-art ASRs for most classes. Additionally, our method demonstrates superior performance on head classes and some tail classes (\textit{e.g.}, class 9) in terms of ACC, as illustrated in Fig.~\ref{fig:classwise-acc-medium}.

\section{Negative Impact and Limitations.}
% \textbf{Negative Impact and Limitations.}
We brings attention to the threat of this practical long-tailed backdoor attack task. While our method achieves a good attack performance, we mainly focus on data augmentation-based methods, and does not consider re-weighting-based techniques. We leave them for an interesting future work.

\end{document}